\documentclass[sigconf]{acmart}

\usepackage{microtype}
\usepackage{rotating}
\usepackage{booktabs}
\usepackage{colortbl}
\usepackage{enumitem}
\usepackage{tabularx}
\usepackage{multirow}
\usepackage{subcaption}
\usepackage{algorithm}
\usepackage{algpseudocode}
\usepackage{pifont}

\usepackage{listings}
\usepackage[most]{tcolorbox}
\tcbuselibrary{breakable}

\definecolor{codegreen}{rgb}{0,0.6,0}
\definecolor{codegray}{rgb}{0.5,0.5,0.5}
\definecolor{codepurple}{rgb}{0.58,0,0.82}
\definecolor{backorange}{RGB}{255,250,240}
\definecolor{frameorange}{RGB}{255,140,0}
\definecolor{codebg}{rgb}{0.95,0.95,0.95}

\newcommand{\eg}{\emph{e.g.}}
\newcommand{\ie}{\emph{i.e.}}

\usepackage[table,xcdraw]{xcolor}
\newcommand{\cmark}{{\color{green!82!black}\ding{51}}}

\newcommand{\xmark}{{\color{red!90!black}\ding{55}}}

\lstdefinestyle{mystyle}{
    backgroundcolor=\color{backorange},   
    commentstyle=\color{codegreen},
    keywordstyle=\color{magenta},
    numberstyle=\tiny\color{codegray},
    stringstyle=\color{codepurple},
    basicstyle=\ttfamily\footnotesize,
    breakatwhitespace=false,         
    breaklines=true,                 
    captionpos=b,                    
    keepspaces=true,                 
    numbers=left,                    
    numbersep=5pt,                  
    showspaces=false,                
    showstringspaces=false,
    showtabs=false,                  
    tabsize=2
}
\newtcolorbox{pythonbox}[1][]{
  enhanced,
  breakable,
  title={Python Code Using COPT},
  colframe=frameorange,
  colback=backorange,
  coltitle=white,
  fonttitle=\bfseries,
  attach boxed title to top left={xshift=5mm, yshift*=-\tcboxedtitleheight/2},
  boxed title style={
    frame hidden, size=small, boxrule=0pt, arc=4pt, colback=frameorange
  },
  arc=3mm, boxrule=0.5pt, top=12pt,
  #1
}

\usepackage{mathtools,amsthm}
\usepackage[capitalize,noabbrev]{cleveref}
\usepackage{adjustbox}

\theoremstyle{plain}

\theoremstyle{definition}

\theoremstyle{remark}

\newcommand{\benchmark}{\textsc{PhotoBench}}

\setcopyright{none}
\renewcommand\footnotetextcopyrightpermission[1]{} 

\begin{document}

\title{PhotoBench: Beyond Visual Matching Towards Personalized Intent-Driven Photo Retrieval}

\author{Tianyi Xu}
\authornote{Equal Contribution.}
\email{crimsonflag@sjtu.edu.cn}
\affiliation{%
  \institution{Shanghai Jiao Tong University}
  \city{Shanghai}
  \country{China}
}

\author{Rong Shan}
\authornotemark[1]
\email{shanrong@sjtu.edu.cn}
\affiliation{%
  \institution{Shanghai Jiao Tong University}
  \city{Shanghai}
  \country{China}
}

\author{Junjie Wu}
\authornotemark[1]
\email{wujunjie1@oppo.com}
\affiliation{%
  \institution{OPPO}
  \city{Shenzhen}
  \country{China}
}

\author{Jiadeng Huang}
\email{huangjiadeng@oppo.com}
\affiliation{%
  \institution{OPPO}
  \city{Shenzhen}
  \country{China}
}

\author{Teng Wang}
\email{wt0318@connect.hku.hk}
\affiliation{%
  \institution{OPPO}
  \city{Shenzhen}
  \country{China}
}

\author{Jiachen Zhu, Wenteng Chen}
\email{{gebro13, cwt-03}@sjtu.edu.cn}
\affiliation{%
  \institution{Shanghai Jiao Tong University}
  \city{Shanghai}
  \country{China}
}

\author{Minxin Tu, Quantao Dou}
\email{{tuminxin, douquantao}@oppo.com}
\affiliation{%
  \institution{OPPO}
  \city{Shenzhen}
  \country{China}
}

\author{Zhaoxiang Wang}
\email{steven.wangzx@gmail.com}
\affiliation{%
  \institution{OPPO}
  \city{Shenzhen}
  \country{China}
}

\author{Changwang Zhang}
\authornote{Corresponding author.}
\email{changwangzhang@foxmail.com}
\affiliation{%
  \institution{OPPO}
  \city{Shenzhen}
  \country{China}
}

\author{Weinan Zhang}
\authornotemark[2]
\email{wnzhang@sjtu.edu.cn}
\affiliation{%
  \institution{Shanghai Jiao Tong University}
  \city{Shanghai}
  \country{China}
}

\author{Jun Wang}
\authornotemark[2]
\email{junwang.lu@gmail.com}
\affiliation{%
  \institution{OPPO}
  \city{Shenzhen}
  \country{China}
}

\author{Jianghao Lin}
\authornotemark[2]
\email{linjianghao@sjtu.edu.cn}
\affiliation{%
  \institution{Shanghai Jiao Tong University}
  \city{Shanghai}
  \country{China}
}

\renewcommand{\shortauthors}{Xu and Shan, et al.}

\begin{abstract}
Personal photo albums are not merely collections of static images but \textit{living, ecological archives} defined by temporal continuity, social entanglement, and rich metadata, which makes the personalized photo retrieval non-trivial. 
However, existing retrieval benchmarks rely heavily on context-isolated web snapshots, failing to capture the \textit{multi-source reasoning} required to resolve authentic, intent-driven user queries. 
To bridge this gap, we introduce \textbf{\benchmark}, the first benchmark constructed from authentic, personal albums. 
It is designed to shift the paradigm from visual matching to personalized multi-source intent-driven reasoning. 
Based on a rigorous \textit{multi-source profiling} framework, which integrates visual semantics, spatial-temporal metadata, social identity, and temporal events for each image, we synthesize complex intent-driven queries rooted in users' life trajectories.
Extensive evaluation on \benchmark{} exposes two critical limitations: the \textbf{modality gap}, where unified embedding models collapse on non-visual constraints, and the \textbf{source fusion paradox}, where agentic systems perform poor tool orchestration. 
These findings indicate that the next frontier in personal multimodal retrieval lies beyond unified embeddings, necessitating robust agentic reasoning systems capable of precise constraint satisfaction and multi-source fusion. Our \benchmark{} is available\footnote{\url{https://github.com/LaVieEnRose365/PhotoBench}}.
\end{abstract}




\maketitle


\section{Introduction}

\begin{table*}[ht]
\centering
\caption{We compare \benchmark{} with existing vision-language or multimodal retrieval datasets, highlighting its unique characteristics for real-world personalized photo retrieval tasks.
\textit{Quality Var.}: Image Quality Variance (\eg, blur, noise); 
\textit{Near-dup.}: Near-duplicate burst shots.
}
\label{tab:intro_compare_table}

\resizebox{0.99\textwidth}{!}{
\renewcommand\arraystretch{1.2}
\begin{tabular}{c|ccccccc|cccccc}
\hline
\multirow{2}{*}{Dataset} & \multicolumn{7}{c|}{Query Dimensions}    & \multicolumn{6}{c}{Image Dimensions} \\ \cline{2-14} 
 & One-to-Many & Unmatched & Narrative & Personalized & Multi-Source & Reasoning & Volume & Personalized & Metadata & Temporal & Quality Var. & Near-dup. & Candidates \\ \hline
MSCOCO\_t2i~\cite{lin2014microsoft}              & \xmark & \xmark & \xmark & \xmark & \xmark & \xmark & 500K & \xmark  & \xmark  & \xmark & \xmark & \xmark & 123K \\
Flickr30k~\cite{plummer2015flickr30k}                & \xmark & \xmark & \xmark & \xmark & \xmark & \xmark & 16K  & \xmark  & \xmark  & \xmark & \xmark & \xmark & 31K  \\
Winoground~\cite{thrush2022winoground}               & \xmark & \xmark & \xmark & \xmark & \xmark & \cmark & 800  & \xmark  & \xmark  & \xmark & \xmark & \xmark & 2    \\
INQUIRE~\cite{vendrow2024inquire}                  & \cmark & \xmark & \xmark & \xmark & \xmark & \cmark & 250  & \xmark  & \cmark  & \cmark & \cmark & \cmark & 5M   \\
VisDial~\cite{das2017visual}                  & \xmark & \xmark & \cmark & \xmark & \cmark & \cmark & 1.2M & \xmark  & \xmark  & \xmark & \xmark & \xmark & 100  \\
VisualNews~\cite{liu2021visual}               & \xmark & \xmark & \cmark & \xmark & \cmark & \cmark & 1.1M & \xmark  & \cmark  & \xmark & \cmark & \cmark & 1M   \\
Wiki-SS-NQ~\cite{ma2024unifying}               & \cmark & \cmark & \xmark & \xmark & \cmark & \cmark & 10K  & \xmark  & \cmark  & \xmark & \cmark & \xmark & 1M   \\ \hline
PhotoBench               & \cmark & \cmark & \cmark & \cmark & \cmark & \cmark & 1.1K+ & \cmark  & \cmark  & \cmark & \cmark & \cmark & 1K \\ \hline
\end{tabular}}
\end{table*}

Personal photo galleries have evolved from static storage into the primary repository of human memory. Unlike curated web-scale datasets (\eg, LAION~\cite{schuhmann2022laion} or MSCOCO~\cite{lin2014microsoft}) where images are isolated snapshots of visual content, a personal album is a \textit{living, ecological archive}. It is temporally continuous, socially entangled, and deeply personalized. 
In this context, user queries are not merely simple visual descriptions (\eg, \textit{a black dog}), but intent-driven requests anchored in heterogeneous signals, such as specific events, social relationships, or spatial-temporal constraints (\eg, \textit{the dinner with my parents before the flight}). Consequently, effective retrieval requires not merely visual matching, but \textit{multi-source reasoning} to fuse visual perception with user-specific context.


Despite the rapid progress in multimodal retrieval, existing benchmarks are generally \textbf{non-personalized} and fail to capture this ecological complexity, as summarized in Table~\ref{tab:intro_compare_table}. 
We identify two critical limitations in current research:
\begin{itemize}[leftmargin=10pt]
    \item \textbf{Lack of Ecological Fidelity (Image Gap)}. Benchmarks like MSCOCO~\cite{lin2014microsoft} or Flickr30k~\cite{plummer2015flickr30k} focus on web-scraped, context-isolated images. They lack the \textit{temporal continuity} and \textit{rich metadata} (timestamps, GPS, identity graphs) inherent to personal albums, rendering them unsuitable for testing temporal- or social-based complex reasoning.
    \item \textbf{Shallow User Intent (Query Gap)}. 
    Datasets such as INQUIRE~\cite{vendrow2024inquire} or VisualNews~\cite{liu2021visual} often rely on descriptive captions that map directly to visual content, which is usually sparse, incomplete one-to-one mapping. 
    They fail to capture the \textit{multi-source entanglement} and \textit{evolving user intent} of real-world queries, where visual signals must be fused with non-visual constraints (\eg, specific time or social role) to resolve the ambiguity.
\end{itemize}

To this end, we introduce \textbf{\benchmark}, a diagnostic benchmark explicitly designed to shift the field of multimodal retrieval from \textit{visual matching} to \textit{personalized intent-driven reasoning}. 
Unlike prior efforts, \benchmark{} is constructed from authentic, personal albums, preserving the natural noise, burstiness, and metadata headers of real-world photography. 
We employ a rigorous \textit{multi-source profiling} to model each photo not merely as pixels, but as an information union of visual semantics $\mathcal{V}$, spatial-temporal metadata $\mathcal{M}$, social identity $\mathcal{F}$, and temporal events $\mathcal{E}$. 
Furthermore, we conduct \textit{intent-driven query synthesis} by conditioning query generation on the multi-source information and the user's life trajectory.
In this way, we reconstruct the latent motivation behind the visual photos.
Finally, through \textit{exhaustive ground truth mining and verification}, we produce complex queries associated with a comprehensive ground truth set, which necessitates cross-modal reasoning in heterogeneous, personalized contexts. 
Moreover, we also synthesize zero-ground-truth queries to evaluate a retrieval system's rejection capability to resist the user's ``false memory''.

Evaluating on SOTA retrieval models and systems, \benchmark{} exposes critical architectural failures that remain hidden in standard benchmarks. Our experiments reveal two critical phenomena:
\begin{itemize}[leftmargin=10pt]
    \item \textbf{Modality Gap}. Unified embedding models (\eg, VLM2Vec~\cite{meng2025vlm2vec} collapse when queries require precise non-visual constraints (Metadata or Face), revealing that they function primarily as visual similarity calculators rather than holistic multi-source reasoners.
    \item \textbf{Source Fusion Paradox}. While agentic retrieval systems equipped with external tools outperform embedding models, they exhibit a non-linear performance degradation as query complexity increases. 
    We find that strong single-source capabilities do not automatically translate to reliable multi-source fusion, highlighting a fundamental bottleneck in \textit{tool orchestration} and \textit{constraint satisfaction} for complex personalized photo retrieval.
\end{itemize}
These findings indicate that the next frontier in personal multimodal retrieval, especially for photo album scenarios, lies \textit{not only} in establishing stronger unified embedding models, \textit{but also} in developing robust and lightweight agentic reasoning systems capable of traversing the modality gap and resolving the source fusion paradox. 
We believe that our \benchmark{} could serve as the key testbed for this evolution.

In summary, our main contributions are:
\begin{itemize}[leftmargin=10pt]
    \item We introduce \textbf{\benchmark}, the first multimodal retrieval benchmark derived from authentic, metadata-rich personal albums. 
    Through \textit{multi-source profiling} for each image, it provides the dense context necessary to evaluate complex reasoning on multi-source personalized information beyond visual matching.
    \item We propose \textbf{intent-driven query synthesis}, which is a \textit{generalized} methodology for personalized multimodal retrieval query generation. 
    It synthesizes narrative yet complex queries rooted in users' life trajectories, followed by exhaustive ground truth mining for comprehensive evaluation.
    Moreover, zero-ground-truth queries are also introduced to evaluate the system reliability.
    \item Experiments on \benchmark{} demonstrate that current retrieval models and systems struggle to fulfill the personalized multi-source photo retrieval task.
    By identifying the \textbf{modality gap} and \textbf{source fusion paradox}, we point out a critical direction for personalized multimodal retrieval, especially for photo album scenarios, \ie, shifting from unified embedding-centric paradigms to robust and lightweight agentic retrieval systems. 
\end{itemize}

\section{Related Work} \label{sec:related}

\textbf{Multimodal Retrieval Benchmarks.} Multimodal retrieval has moved beyond object matching toward semantic understanding. Early benchmarks like MSCOCO~\cite{lin2014microsoft} and Flickr30k~\cite{plummer2015flickr30k} focused on simple matching between visual content and captions. Recent datasets introduce more complexity: Winoground~\cite{thrush2022winoground} tests compositional reasoning, while INQUIRE~\cite{vendrow2024inquire} evaluates retrieval in large-scale pools. Other works target specific domains, such as Visual News~\cite{liu2021visual} for metadata, VisDial~\cite{das2017visual} for conversational search, and LSC~\cite{gurrin2023introduction} for lifelogging trajectories. Composed Image Retrieval (CIR) datasets, including Fashion IQ~\cite{fashioniq} and CIRR~\cite{cirr}, further expand the field by using a reference image with text modifications. However, existing benchmarks often focus on visual content rather than the user's underlying intent. \benchmark{} addresses this by mapping complex queries to user motivations in personal photo albums, covering continuous life segments and long-term memory.

\textbf{Multimodal Representation.} Retrieval performance depends on how well models represent data in a latent space. CLIP~\cite{radford2021learning} and SigLIP~\cite{zhai2023sigmoid} established the foundation for cross-modal alignment. To improve discriminative power, recent methods use hard negative gradient amplification~\cite{xue2025improvemultimodalembeddinglearning} and smart batch mining~\cite{thirukovalluru2025breaking}. With the rise of generative multimodal large language models (MLLMs), recent works start to train embedding models using MLLMs as backbones. Models like Qwen3-VL~\cite{li2026qwen3, yang2025qwen3} and InternVL~\cite{chen2024internvl, lu2025internvl} are now used as powerful encoders through frameworks like VLM2Vec~\cite{jiang2024vlm2vec, meng2025vlm2vec} and Rzenembed~\cite{jian2025rzenembed}. Mapping-based methods, such as FiRE~\cite{fire} and iSEARLE~\cite{searle}, also use MLLMs for textual inversion and context learning~\cite{lincir, cireVL}. Despite these gains, these embeddings in a unified latent semantic space often fail to decompose personal queries that involve complex metadata or OCR results~\cite{chen2024bge}.

\textbf{Agentic and Reasoning-based Retrieval.} As tasks grow more complex, researchers are moving from static matching toward reasoning-based search. Following frameworks like ReAct~\cite{yao2022react} and Reflexion~\cite{shinn2023reflexion}, agentic retrieval systems solve tasks by calling external tools. This trend includes MMSearch~\cite{jiang2024mmsearchbenchmarkingpotentiallarge}, AutoCIR~\cite{autocir} and XR~\cite{xr}, which use multi-agent collaboration for retrieval, and MRA-CIR~\cite{MRACIR}, MM-R1~\cite{liang2025mmr1unleashingpowerunified}, MMSearch-R1~\cite{wu2025mmsearchr1incentivizinglmmssearch} which employs reasoning agents for open-ended search. Other strategies, such as LDRE~\cite{ldre} and \textit{Reason-before-retrieve}~\cite{Reason-before-retrieve}, use divergent thinking to clarify ambiguous queries before searching. While these approaches represent the current frontier, they need challenging benchmarks to test their reliability and ability to handle unanswerable questions~\cite{kirichenko2025abstentionbench}. \benchmark{} provides a unified framework to evaluate both traditional embedding models and newer agentic pipelines.

\section{Dataset Construction}
\label{dataset_construction}

\begin{figure*}[ht]
    \centering
    \includegraphics[width=0.97\textwidth]{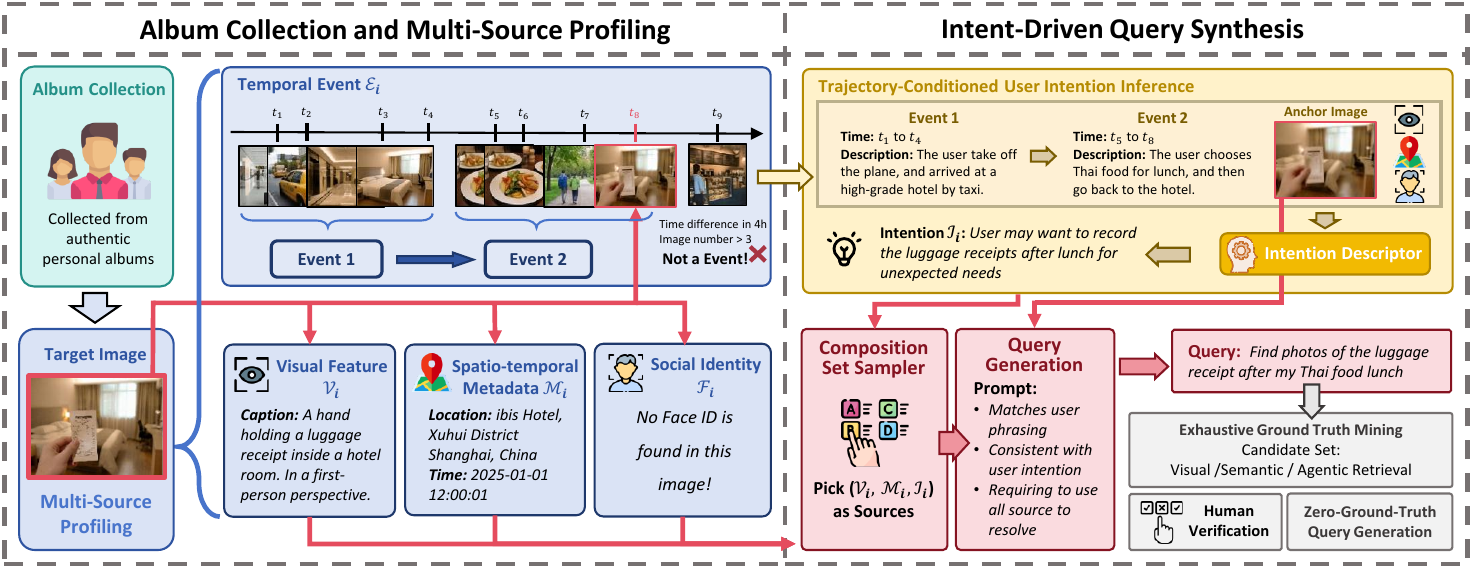}
    \caption{The illustration of the dataset construction pipeline for \benchmark{}.}
    \label{fig:pipeline}
\end{figure*}

Unlike web-scale image collections characterized by isolated snapshots, photo albums are defined by their \textbf{ecological validity}: (1) the images are temporally continuous, socially entangled, and deeply personalized, and (2) the user queries are intent-driven and depend on multi-source private contexts.
\textsc{PhotoBench} is explicitly designed to capture this contextual depth, distinguishing it from general-purpose multimodal retrieval benchmarks. 
The construction process is divided into two stages:
\begin{itemize}[leftmargin=10pt]
    \item \textbf{Album collection and multi-source profiling.} We collect authentic, temporally continuous personal albums and construct a structured, multi-source profile for each image. 
    This process integrates fine-grained visual semantics, high-fidelity spatio-temporal metadata, social cues, and hierarchical event structures.
    \item \textbf{Intent-driven query synthesis.} Rather than relying on static descriptive captions, we synthesize queries by inferring user intentions from their life trajectory. 
    This is followed by a rigorous, expert-verified mining process to ensure exhaustive ground truth recall, alongside the generation of diagnostic zero-ground-truth queries to test system reliability (\ie, rejection capability).
\end{itemize}
Due to page limitation, we provide all the prompts used for LLM or MLLM generation in Appendix~\ref{app:prompts}. 


\subsection{Album Collection \& Multi-Source Profiling}

\subsubsection{Album Collection}
\label{sec:album_collection}

To ensure high ecological validity, \textsc{PhotoBench} is constructed from authentic, temporally continuous photo albums acquired from a diverse demographic pool, spanning varied age groups and professional backgrounds. 
The collection protocol adheres to two governing policies:
\begin{itemize}[leftmargin=10pt]
    \item \textbf{Holistic metadata retention.} We posit that photo album retrieval is inextricably linked to non-visual context. Consequently, we prioritize albums with high metadata integrity, ensuring that all collected assets retain their original high-fidelity spatio-temporal signals (\eg, timestamps, GPS coordinates, and device headers) exactly as captured by the source devices.
    
    \item \textbf{Privacy review with minimal curation.} We purchased complete albums from consenting participants, rather than selectively sampling representative photos. To make \textsc{PhotoBench} publicly releasable, we conducted a privacy screening process that combines user feedback and expert review. Participants were allowed to flag sensitive content, and experts further removed or masked images containing highly private or identifying information (\eg, confidential documents, personal IDs or other content unsuitable for public release). Beyond this necessary privacy filtering, we intentionally avoided additional manual pruning or aesthetic curation, so that the albums maintain their original characteristics and natural photo distributions.
\end{itemize}




\subsubsection{Multi-Source Profiling}
\label{sec:image_profiling}

Each collected photo inherently provides its visual content and on-device metadata. 
Therefore, for each image $i$, we begin profiling from two native sources, \ie, visual feature and metadata as follows:
\begin{itemize}[leftmargin=10pt]
    \item \textbf{Visual features \(\mathcal{V}_i\)}. Beyond the raw image itself, we use an MLLM (\ie, GPT-4o) to extract and caption the fine-grained visual semantics that are frequently referenced in real queries, including salient objects, human poses, scene composition, and aesthetic attributes.
    \item \textbf{Spatio-temporal metadata \(\mathcal{M}_i\)}. We transform low-level metadata into semantic descriptors. Raw GPS coordinates are mapped to semantic places-of-interest (POIs) via reverse geocoding, and timestamps are normalized into human-like temporal tags, \eg, early morning, weekend, and Halloween.
\end{itemize}
Beyond intrinsic attributes \(\mathcal{V}_i\) and \(\mathcal{M}_i\), personal albums are structured by social relations and temporal narratives. We therefore augment the profile with two relational components:
\begin{itemize}[leftmargin=10pt]
    \item \textbf{Social identity \(\mathcal{F}_i\).} We construct a local social graph for each album via face detection and clustering. Following ego-identification (isolating the album owner), human experts annotate recurring identity clusters with plausible social roles (\eg, spouse, colleague) based on co-occurrence patterns. This produces structured and personalized cues that enable relational references.
    \item \textbf{Temporal event \(\mathcal{E}_i\).} 
    To reconstruct the user's life trajectory, we perform hierarchical temporal clustering. Photos exhibiting temporal proximity within a definable window form an event cluster. In our implementation, we utilize an adaptive 4-hour window with a minimum cluster size of 3 images. Each cluster is assigned a concise textual summary (\eg, \textit{business dinner at a Japanese restaurant}), providing the necessary temporal context for the subsequent inference of user intention.
\end{itemize}
With these steps, each image $i$ is associated with a temporally dependent and multi-source profile, represented as:
\begin{equation}
\label{eq:image_profile}
\mathcal{P}_i = \{\mathcal{V}_i, \mathcal{M}_i, \mathcal{F}_i, \mathcal{E}_i\},
\end{equation}
which serves as the structured data foundation for the subsequent intent-driven query synthesis stage.

\subsection{Intent-Driven Query Synthesis}
\label{sec:intent_query_synthesis}

Authentic retrieval requests in personal archives are rarely simple visual descriptions; they are anchored in specific memory traces and grounded in heterogeneous, personalized contexts. To emulate this complexity, we employ an intent-driven query synthesis pipeline. 
Specifically, we construct each query by starting with an anchor image \(i\) and its profile \(\mathcal{P}_i\). We first infer the \textit{user intention} behind the image by analyzing the user's event trajectory, and then synthesize a natural query by composing information from multiple profile dimensions. 
Crucially, to ensure robust evaluation, we conduct exhaustive ground truth mining and expert verification to establish dense ground truth.
Furthermore, we also generate some zero-ground-truth queries that have no ground truth images to evaluate the system's rejection capability.


\subsubsection{Trajectory-Conditioned User Intention Inference}
\label{sec:intention_inference}

A single photo is often only a snapshot of a broader activity, and its purpose is best understood in the context of evolving events. 
For each image \(i\), we infer an intention descriptor \(\mathcal{I}_i\) by conditioning on its profile \(P_i\) and the textual summaries of preceding events:
\begin{equation}
\label{eq:intent_def}
\mathcal{I}_i = \operatorname{MLLM}\big(\mathcal{P}_i,\ [\mathcal{E}_j]_{j \le i}\big),
\end{equation}
where \([\mathcal{E}_j]_{j\le i}\) denotes the chronological trajectory of event summaries that occur earlier in the album timeline. The output \(\mathcal{I}_i\) is a concise, human-like description of the user’s potential intention (\eg, \textit{to keep the dinner receipt during a business trip}). 
In this way, we encourage the model to exploit trajectory context rather than relying on single-image captioning.

\subsubsection{Query Synthesis via Multi-source Composition}
\label{sec:query_composition}

Given the inferred intention \(\mathcal{I}_i\) and the profile $\mathcal{P}_i$, we synthesize diverse queries for image $i$ by composing multiple available information sources. Specifically, we sample a composition set:
\begin{equation}
\mathcal{H} \subseteq \{\mathcal{V}_i, \mathcal{M}_i, \mathcal{F}_i, \mathcal{I}_i\},
\end{equation}
and prompt an LLM/MLLM to generate a natural-language query \(q\) that: (1) mirrors realistic user phrasing, (2) remains logically consistent with \(\mathcal{I}_i\), and (3) strictly requires the intersection of sources in \(\mathcal{H}\) to resolve ambiguity in a dense gallery. This mechanism yields narrative, personalized queries that closely approximate human memory retrieval patterns (see Appendix~\ref{app:query_case_study} for examples).

\subsubsection{Exhaustive Ground Truth Mining and Verification}
\label{sec:candidate_mining_verification}

In ecological photo retrieval, a single query often corresponds to multiple valid targets (\eg, burst shots, near-duplicates, or thematically linked events) among numerous hard distractors. 
To ensure \textsc{PhotoBench} supports rigorous recall evaluation, we move beyond the single anchor image \(i\) and perform \textit{exhaustive ground truth mining} for the synthesized query $q$.
Concretely, we construct a candidate set by combining complementary retrieval methods:
\begin{itemize}[leftmargin=10pt]
    \item \textbf{Visual retrieval:} Top-\(K\) neighbors based on visual embedding similarity, capturing perceptually identical near-duplicates.
    \item \textbf{Semantic retrieval:} Top-\(K\) neighbors using text embeddings over query $q$ and caption from $\mathcal{P}_i$, to capture the semantically relevant but visually distinct photos.
    \item \textbf{Agentic multi-tool retrieval:} A rigorous agentic pipeline that filters the album using metadata, identity, and event constraints to uncover valid matches that embedding models might miss.
\end{itemize}
Empirically, we set \(K=50\), which effectively covers the valid solution space. Finally, human experts manually review every instance in the candidate pool to annotate all valid positives and filter out ambiguous or ill-posed queries. 
This combination of automated exhaustive mining and human verification ensures that our \textsc{PhotoBench} provides a \textbf{comprehensive ground truth set}, distinguishing it from web-scale datasets where labels are often sparse and incomplete.

\subsubsection{Zero-Ground-Truth Query Generation}
\label{sec:zero_gt_queries}

To evaluate a retrieval system's rejection capability to resist user hallucination, we simulate ``false memory'' scenarios where users query for plausible but non-existent images (\eg, a specific person at an event they did not attend). We generate a set of \textit{zero-ground-truth (Zero-GT)} queries using counterfactual synthesis. Detailed generation protocols are provided in Appendix~\ref{app:zero_gt_construct}.

\section{Dataset Statistics}
\label{sec:statistics}

In this section, we present a statistical overview of \textsc{PhotoBench}, focusing on its overall characteristics and our proposed source-aware taxonomy. 
We also provide the detailed dataset statistics and distribution analysis in Appendix~\ref{app:stats}.

\subsection{Dataset Overview}

Following the ecological construction protocol outlined in Section~\ref{dataset_construction}, \benchmark{} consists of 3,582 images sampled from three authentic, personal albums, paired with 1,188 bilingual queries (Chinese and English). Unlike web-scraped datasets characterized by sparse labels, \benchmark{} offers \textit{dense, exhaustive ground truth} within a continuous life-logging context. 
It exhibits three critical characteristics designed to test complex retrieval capabilities:

\begin{itemize}[leftmargin=10pt]
    \item \textbf{Spatio-Temporal Fidelity:} 83.4\% of images retain valid high-precision GPS and timestamp metadata. 
    The data covers a temporal range from 2018 to 2025, capturing diverse settings from major metropolitan areas in China to international POIs across East and Southeast Asia. 
    This is essential for evaluating the retrieval systems' ability to perform rigorous spatio-temporal filtering, which is often untested in metadata-stripped web datasets.
    \item \textbf{Social-Relational Density:} The albums collectively feature 20 distinct recurring individuals, with 25.1\% of images classified as portraits. This density enables the evaluation of long-tail social identity recognition and relationship-based retrieval.
\end{itemize}
We provide per-album statistical analysis in Appendix~\ref{app:comp_benches}.

\subsection{Source-Aware Query Taxonomy}
\label{sec:query_taxonomy}

To enable precise failure attribution (\eg, distinguishing visual failures from reasoning failures), we introduce a \emph{Source-Aware Query Taxonomy}. Curated via expert annotation, this taxonomy classifies queries based on the \textit{necessary and sufficient} information sources required to resolve the user query for image retrieval. We define three atomic information dimensions:
\begin{itemize}[leftmargin=10pt]
    \item \textbf{Vision (${S}_{V}$):} Queries resolvable solely through visual perception, involving physical entities, scene aesthetics, or composition (\eg, \textit{photo of red flowers}).
    \item \textbf{Metadata (${S}_{M}$):} Queries grounded in spatio-temporal context, requiring access to timestamp or geolocation logs (\eg, \textit{photos from Tokyo in 2025}).
    \item \textbf{Face (${S}_{F}$):} Queries targeting specific social identities or relationships, requiring face recognition or social graph access (\eg, \textit{photo of my sister}).
\end{itemize}
Real-world intents are often entangled. Queries requiring multiple information sources are classified into \textbf{compositional categories} (\ie, ${S}_{VM}$, ${S}_{VF}$, ${S}_{MF}$, and ${S}_{VMF}$). For instance, ${S}_{VM}$ denotes a query requiring both visual recognition and metadata filtering.
Crucially, this classification is strict and non-overlapping. A query assigned to a composite category (\eg, ${S}_{VM}$) is \textit{not} double-counted under ${S}_{V}$ or ${S}_{M}$. 
This mutual exclusivity provides a rigorous foundation for evaluating retrieval systems’ ability to process and fuse multiple information sources. 

\begin{figure}[t]
    \centering
    \includegraphics[width=0.99\linewidth]{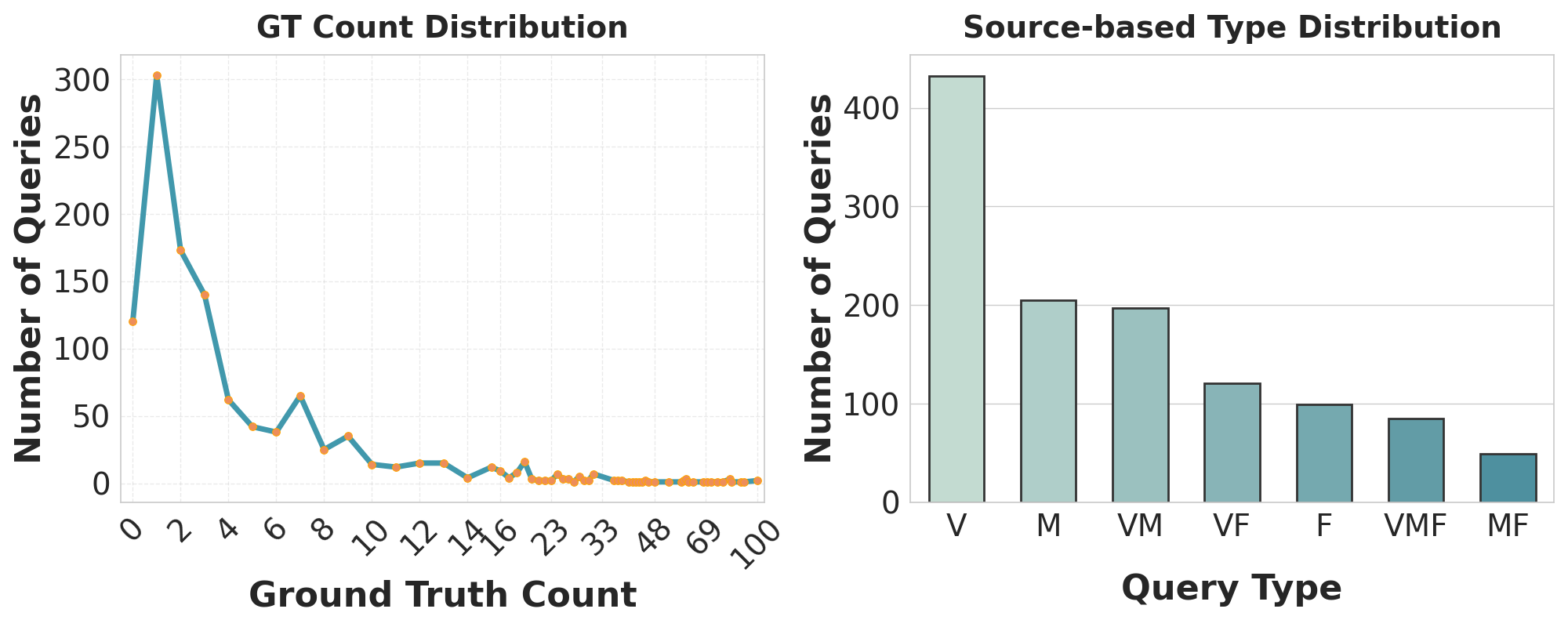}
    \caption{The query distribution w.r.t. ground-truth count (left) and source-aware taxonomy (right) in \benchmark{}.}
    \label{fig:query_stats_overall}
\end{figure}

Figure~\ref{fig:query_stats_overall} presents the query distribution in terms of both ground-truth count and source-aware taxonomy. 
The ground truth distribution exhibits a long-tail characteristic, challenging systems to handle both specific needle-in-a-haystack retrieval and broader event-level recall. 
Notably, a significant proportion of queries fall into composite categories (\eg, ${S}_{VM}$, ${S}_{VMF}$), highlighting that the primary challenge in personal retrieval is not merely visual matching, but the \textit{cross-modal fusion} of heterogeneous signals.





\section{Experiment}

\textsc{PhotoBench} evaluates retrieval systems in a personalized, intent-driven context that differs substantially from static, context-isolated web benchmarks. 
In this section, we benchmark a diverse spectrum of retrieval paradigms to establish rigorous baselines for the research community. 
We report primary results on the Chinese version of \benchmark{}. 
We provide the case studies of various failure modes in \benchmark{} in Appendix~\ref{app:cases} due to the page limitation.

\subsection{Evaluated Models and Systems}
\label{sec:evaluated_methods}

We evaluate two distinct architectural families: \textit{Unified Embedding Models} (single-step matching) and \textit{Hybrid Retrieval Systems} (multi-step, tool-augmented reasoning).

\subsubsection{Unified Embedding Models}
These methods map images and queries into a shared latent space for similarity-based retrieval. We categorize them into two subgroups:
\begin{itemize}[leftmargin=10pt]
    \item \textbf{Multimodal embedding models} that directly embed visual and textual inputs. 
    We select representative baselines across varying scales, including dual-encoder models (\ie, CLIP~\cite{radford2021learning}, SigLIP2-base/giant~\cite{tschannen2025siglip}) and VLM-based dense retrievers (\ie, VLM2Vec~\cite{jiang2024vlm2vec}, B3\_Qwen2\_7B~\cite{thirukovalluru2025breakingbatchbarrierb3}, Qwen3-VL-Embedding-2B/8B~\cite{li2026qwen3}, Ops-MM-embedding-v1~\cite{Ops-MM}, RzenEmbed-v2-7B~\cite{jian2025rzenembed} and QQMM-embed-v2~\cite{xue2025improvemultimodalembeddinglearning}).
    \item \textbf{Caption-based text embedding pipelines} that first convert images to textual captions using GPT-4o and then perform text-to-text retrieval. 
    For the textual encoder, we employ BGE-M3~\cite{chen2024bge}, multilingual-E5 series~\cite{wang2024multilingual}, and Qwen3-Embedding series~\cite{zhang2025qwen3embeddingadvancingtext}.
\end{itemize}

\subsubsection{Hybrid Retrieval Systems}
These systems utilize compositional reasoning or heuristic logic to combine multiple retrievers and signals (\eg, visual similarity, metadata constraints), typically returning a \textit{variable-length} result set.
\begin{itemize}[leftmargin=10pt]
    \item \textbf{Tool-based agentic systems}.
    To evaluate the potential of LLM in resolving complex retrieval intents, we implement a \textit{ReAct-style agent}~\cite{yao2022react} instantiated with various SOTA LLM backbones capable of tool invocation, including ToolACE-2-Llama-3.1-8B~\cite{liu2025toolace}, Qwen3-8B/32B~\cite{yang2025qwen3}, DeepSeek-V3~\cite{deepseekai2025deepseekv3technicalreport}, Qwen3-235B-A22B~\cite{yang2025qwen3}, GPT-4o~\cite{openai2024gpt4ocard}, OpenAI-o3~\cite{OpenAI-o3}, Claude-Sonnet-4-5~\cite{claude-sonnet}, and Claude-Opus-4-5~\cite{claude-opus}. 
    To strictly align with the information sources defined in Section~\ref{sec:query_taxonomy}, we equip the agent with following tools:
    \begin{itemize}[leftmargin=10pt]
        \item \textbf{Vector search engine (${T}_V$)} performs semantic visual retrieval over a FAISS-indexed embedding space. 
        We adopt RzenEmbed-v2-7B~\cite{jian2025rzenembed} as the encoder, as it demonstrates superior performance among unified embedding models (see Section~\ref{sec:main_results}).
        \item \textbf{Metadata filter (${T}_M$)} executes hard filtering based on spatio-temporal constraints (timestamps, GPS coordinates, and POIs).
        \item \textbf{Face search engine (${T}_F$)} resolves identity constraints and social references using face clusters and annotated role tags.
        \item \textbf{Set composition tool ($T_S$)} enables the agent to perform logical set operations (intersection/union/difference) over outputs from different tools to synthesize the final prediction.
    \end{itemize}

    \item \textbf{Real-world mobile gallery systems.} 
    To contextualize academic findings against industrial standards, we evaluate the native gallery search performance of six mainstream flagship smartphones, covering the three dominant ecosystems: iOS, Android and HarmonyOS.
    We strictly follow a black-box evaluation protocol: albums are imported into the native gallery, indexed by the on-device engine, and queried via an automated scripting tool. 
    To focus the evaluation on the representative capabilities of mature commercial systems rather than the specific performance of individual brands, we have anonymized the devices as \texttt{Phone A} to \texttt{Phone F}. The detailed evaluation protocol is in Appendix~\ref{app:phone_detail}
\end{itemize}

\subsection{Evaluation Metrics}
\label{sec:metrics}

\benchmark{} presents two evaluation challenges: (1) it supports one-to-many matches with variable ground truth sizes, and (2) it includes zero-ground-truth (Zero-GT) queries that require system abstention. 
Hence, we employ two complementary metric families:
\begin{itemize}[leftmargin=10pt]
    \item \textbf{Top-\(K\) Ranking Metrics.}
    Designed for embedding models that output \textit{fixed-length lists}. We report Recall@K and NDCG@K with \(K \in \{1, 5, 10, 20\}\), covering the spectrum from best hit to broad shortlists.
    
    \item \textbf{Set-Based Metrics.}
    Only suitable for hybrid retrieval systems (\ie, Agents and Phones) that return \textit{variable-length sets}. 
    We evaluate performance across two query types:
    \begin{itemize}[leftmargin=10pt]
        \item \textbf{Normal Query}. 
        We report standard \texttt{Precision}, \texttt{Recall}, and \texttt{F1} to measure the accuracy of the returned image set against the comprehensive ground truth set.
        \item \textbf{Zero-GT Query.} To measure systems' ability to correctly abstain (reject) when no relevant photo exists, we report \texttt{Reject-Precision}, \texttt{Reject-Recall}, and \texttt{Reject-F1}. Here, \texttt{Reject-Recall} measures the proportion of Zero-GT queries correctly identified as empty, while \texttt{Reject-Precision} measures the reliability of an empty response. Formal definitions are provided in Appendix~\ref{appendix: rejection metrics}.
    \end{itemize}
\end{itemize}

\subsection{Main Results}
\label{sec:main_results}

\begin{table*}[t]

\centering
\caption{The retrieval performance on \benchmark{} using top-$k$ ranking metrics. We report Recall@K (R@K) and NDCG@K (N@K) averaged over all albums. Best result of each category is in bold, and the second best is underlined. 
}
\label{tab:main_table}
\resizebox{0.99\textwidth}{!}{
\renewcommand\arraystretch{1.05}
\begin{tabular}{cc|cccccccc|cccccccc}
\toprule
\hline
                              &  & \multicolumn{8}{c|}{Chinese} & \multicolumn{8}{c}{English} \\
\multirow{-2}{*}{Model} &
  \multirow{-2}{*}{\begin{tabular}[c]{@{}c@{}}Size\\ (B)\end{tabular}} &
  R@1 &
  R@5 &
  R@10 &
  R@20 &
  N@1 &
  N@5 &
  N@10 &
  N@20 &
  R@1 &
  R@5 &
  R@10 &
  R@20 &
  N@1 &
  N@5 &
  N@10 &
  N@20 \\ \hline
\multicolumn{18}{c}{\cellcolor[HTML]{EFEFEF}\textit{Multimodal Embedding Models}}                   \\
clip-ViT-B-32-multilingual-v1 & 0.14 & 1.2 & 4.1 & 6.1 & 8.8 & 3.2 & 4.0 & 4.7 & 5.5 & 10.7 & 25.5 & 33.1 & 41.3 & 24.5 & 25.6 & 27.9 & 30.3 \\
siglip2-base-patch16-224      & 0.38 & 16.4 & 33.7 & 40.0 & 47.3 & 33.1 & 34.6 & 36.0 & 38.2 & 19.8 & 39.9 & 46.7 & 53.1 & 40.2 & 41.5 & 42.9 & 44.7 \\
siglip2-giant-opt-patch16-256 & 1.88 & 20.8 & 40.5 & 47.6 & 54.5 & 42.0 & 42.4 & 44.1 & 46.0 & \underline{23.8} & 45.1 & 50.7 & 56.9 & 47.5 & 48.1 & 48.8 & 50.5 \\
VLM2Vec                       & 8.29 & 23.1 & 44.4 & 52.4 & 60.0 & 46.0 & 47.0 & 48.9 & 51.3 & 22.0 & 43.3 & 51.1 & 58.1 & 43.0 & 44.9 & 46.9 & 49.1 \\
B3\_Qwen2\_7B                 & 8.29 & 20.9 & 41.4 & 49.9 & 57.1 & 41.9 & 43.6 & 45.6 & 47.8 & 21.1 & 42.4 & 50.1 & 55.9 & 42.3 & 43.7 & 45.7 & 47.4 \\
Qwen3-VL-Embedding-2B         & 2.13 & 22.7 & 42.5 & 50.6 & 58.2 & 44.6 & 45.4 & 47.4 & 49.5 & 21.6 & 40.3 & 48.1 & 55.9 & 42.9 & 43.2 & 45.1 & 47.6 \\
Qwen3-VL-Embedding-8B         & 8.14 & 24.9 & 46.2 & 53.0 & 59.2 & 49.7 & 49.7 & 50.9 & 52.6 & \underline{23.8} & 45.2 & 51.9 & 58.4 & 46.8 & 47.9 & 49.2 & 50.9 \\
Ops-MM-embedding-v1           & 8.29 & 25.8 & 48.7 & 56.6 & 63.7 & 49.8 & 51.7 & 53.5 & 55.5 & 25.5 & \underline{48.3} & \underline{56.0} & \underline{63.4} & \textbf{50.9} & \textbf{51.8} & \underline{53.2} & \underline{55.3} \\
RzenEmbed-v2-7B               & 8.29 & \textbf{27.2} & \underline{49.9} & \textbf{58.0} & \underline{65.1} & \textbf{54.1} & \textbf{54.3} & \textbf{56.0} & \textbf{57.9} & \textbf{25.6} & \textbf{48.5} & \textbf{56.5} & \textbf{64.1} & \underline{50.8} & \textbf{51.8} & \textbf{53.7} & \textbf{55.8} \\
QQMM-embed-v2                 & 8.29 & \underline{26.6} & \textbf{50.0} & \underline{57.8} & \textbf{65.4} & \underline{52.3} & \underline{53.7} & \underline{55.4} & \underline{57.6} & 25.1 & 47.8 & 55.9 & 63.6 & 49.3 & 50.6 & 52.6 & 54.9 \\ \hline
\multicolumn{18}{c}{\cellcolor[HTML]{EFEFEF}\textit{Caption-based Text Embedding Pipelines}}           \\
multilingual-e5-small         & 0.12 & 21.6 & 37.5 & 42.8 & 48.8 & 40.7 & 39.6 & 40.7 & 42.5 & 19.3 & 35.6 & 41.6 & 47.5 & 37.1 & 37.1 & 38.3 & 40.0 \\
multilingual-e5-base          & 0.28 & 21.8 & 38.1 & 44.8 & 51.9 & 41.6 & 40.5 & 42.2 & 44.3 & 20.8 & 37.7 & 44.3 & 50.4 & 39.0 & 39.3 & 41.0 & 42.8 \\
multilingual-e5-large         & 0.56 & 24.2 & 40.6 & 47.1 & 54.6 & 46.3 & 44.7 & 45.9 & 48.0 & 20.4 & 38.1 & 45.4 & 50.8 & 39.7 & 40.2 & 42.1 & 43.5 \\
bge-m3                        & 0.57 & 23.0 & 41.6 & 48.5 & 56.6 & 43.8 & 44.5 & 46.0 & 48.5 & 20.8 & 38.1 & 44.7 & 51.5 & 41.9 & 40.6 & 42.0 & 43.8 \\
Qwen3-Embedding-0.6B          & 0.60 & 24.0 & 42.0 & 48.4 & 54.9 & 45.9 & 45.5 & 46.8 & \underline{48.6} & 22.2 & 39.3 & 46.1 & 52.5 & 42.0 & 41.7 & 43.1 & 45.0 \\
Qwen3-Embedding-4B            & 4.00 & \textbf{25.6} & \underline{44.5} & \underline{51.6} & \textbf{57.9} & \textbf{49.1} & \textbf{48.3} & \textbf{49.8} & \textbf{51.4} & \underline{24.0} & \underline{42.3} & \underline{49.2} & \textbf{55.0} & \textbf{46.1} & \underline{45.2} & \underline{46.7} & \underline{48.2} \\
Qwen3-Embedding-8B            & 7.60 & \underline{25.4} & \textbf{44.8} & \textbf{51.7} & \underline{57.8} & \underline{48.3} & \underline{48.2} & \underline{49.7} & 51.2 & \textbf{24.3} & \textbf{43.0} & \textbf{49.2} & \underline{54.8} & \underline{46.0} & \textbf{45.9} & \textbf{47.0} & \textbf{48.5} \\ \hline
\multicolumn{18}{c}{\cellcolor[HTML]{EFEFEF}\textit{Tool-based Agentic Systems}}                                  \\
ToolACE-2-Llama-3.1-8B        & 8.00 & 21.3 & 44.0 & 47.7 & 51.2 & 48.8 & 50.9 & 50.2 & 50.7 & 23.0 & 43.5 & 46.4 & 48.2 & 49.4 & 49.4 & 47.7 & 47.4 \\
Qwen3-8B                      & 8.00 & 14.5 & 51.0 & 60.4 & 64.1 & 35.0 & 53.5 & 56.4 & 56.4 & 14.2 & 51.2 & 62.7 & 66.0 & 34.1 & 55.7 & 58.8 & 59.3 \\
Qwen3-32B                     & 32.8 & 28.3 & 54.1 & 62.5 & 64.5 & 63.6 & 63.0 & 63.5 & 62.4 & 27.2 & 54.5 & \underline{62.7} & 66.3 & 59.6 & 61.5 & 61.9 & 62.0 \\
DeepSeek-v3                   & 671 & 26.8 & 51.6 & 60.1 & 62.4 & 59.6 & 59.9 & 61.0 & 60.6 & 24.9 & 49.4 & 57.8 & 60.9 & 55.5 & 56.5 & 57.7 & 57.8 \\
Qwen3-235B-A22B               & 235 & 31.0 & \textbf{60.8} & 67.0 & 69.6 & \underline{70.0} & \textbf{71.4} & \textbf{72.0} & \textbf{71.8} & 27.7 & 55.9 & 60.9 & 62.8 & 63.6 & 65.6 & 65.7 & 65.2 \\
GPT-4o                        & Closed & 28.0 & 53.1 & 58.5 & 60.3 & 60.3 & 61.1 & 60.5 & 59.5 & 27.6 & 54.3 & 60.5 & 62.3 & 60.9 & 61.9 & 61.7 & 61.0 \\
OpenAI-o3                     & Closed & \underline{31.2} & 59.5 & \underline{68.7} & \underline{71.4} & \textbf{70.2} & \underline{69.8} & \underline{70.6} & \underline{69.9} & \underline{29.2} & \underline{58.5} & \textbf{68.2} & \textbf{71.1} & \underline{67.2} & \textbf{67.6} & \textbf{68.4} & \textbf{67.9} \\
Claude-Sonnet-4-5             & Closed & 30.5 & \underline{60.4} & \textbf{69.5} & \textbf{73.1} & 69.0 & 69.7 & 70.3 & 70.0 & \textbf{30.4} & \textbf{58.9} & \underline{68.1} & \underline{70.4} & \textbf{67.3} & \underline{67.5} & \underline{68.3} & \underline{67.6} \\
Claude-Opus-4-5               & Closed & \textbf{32.1} & 60.6 & \underline{68.7} & 71.7 & 69.8 & 69.9 & 70.3 & \underline{69.9} & 28.7 & 58.1 & 66.7 & 69.0 & 65.2 & 66.7 & 67.3 & 66.5 \\ \hline
   \bottomrule
\end{tabular}}
\end{table*}

\begin{table}[t]

\centering
\caption{
The retrieval performance on \benchmark{} using set-based  metrics. 
Best result of each category is in bold, while the second best is underlined.
\textit{P}, \textit{R}, \textit{Rej-P}, \textit{Rej-R} and \textit{Rej-F1} denote \textit{Precision}, \textit{Recall}, \textit{F1}, \textit{Reject-Precision}, \textit{Reject-Recall} and \textit{Reject-F1}, respectively.}
\label{tab:phone_metrics}
\vspace{-8pt}
\footnotesize
\resizebox{0.47\textwidth}{!}{
\renewcommand\arraystretch{1.1}
\begin{tabular}{c|ccc|ccc}
\toprule
\hline
\multirow{2}{*}{System} & \multicolumn{3}{c|}{Normal Query} & \multicolumn{3}{c}{Zero-GT Query} \\
                        & P   & R  & F1  & Rej-P   & Rej-R  & Rej-F1  \\ \hline
\multicolumn{7}{c}{\cellcolor[HTML]{EFEFEF}\textit{Real-world Mobile Gallery Systems}}                                  \\
                        
Phone A                 & 22.8        & 47.3    & 25.6  & 18.2    & 73.2    & 29.1    \\
Phone B                 & 16.0        & 17.2    & 14.4  & 12.4    & \underline{80.5}   & 21.5    \\
Phone C                 & 33.7        & \underline{50.1}    & 34.4  & 20.1    & 48.8    & 28.5    \\
Phone D                 & 34.9        & 47.7    & 34.5  & \underline{25.5}    & 79.7    & \underline{38.6}    \\
Phone E                 & \textbf{40.9}        & 43.3    & \underline{38.3}  & 20.9    & \textbf{82.3}   & 33.4    \\
Phone F                 & \underline{37.6}        & \textbf{65.4}    & \textbf{40.2}  & \textbf{36.5}    & 63.4    & \textbf{46.3}    \\ \hline
\multicolumn{7}{c}{\cellcolor[HTML]{EFEFEF}\textit{Tool-based Agentic Systems}}                                  \\
ToolACE-2               & 38.3        & 54.7    & 39.0  & 15.1    & \textbf{35.0}    & \underline{21.1}    \\
Qwen3-8B                & 37.6        & 68.1    & 41.0  & 16.8    & 16.7    & 16.7    \\
Qwen3-32B               & 39.6        & 66.3    & 41.4  & 16.7    & 25.8    & 20.3    \\
DeepSeek-v3             & 38.5        & 66.2    & 40.7  & 11.8    & 10.0    & 10.8    \\
Qwen3-235B-A22B         & \textbf{50.8} & 71.6 & \textbf{51.9} & 14.3 & 6.7 & 9.1 \\
GPT-4o                  & 45.4        & 65.3    & 46.4  & 13.8    & 17.5    & 15.4    \\
OpenAI-o3               & 44.6        & \underline{75.2}    & 48.0  & \textbf{34.5}    & \underline{33.3}    & \textbf{33.9}    \\
Claude-Sonnet-4.5       & 43.7        & \textbf{78.3}    & 47.9  & 14.3    & 8.3     & 10.5    \\
Claude-Opus-4.5         & \underline{46.9}        & 74.7    & \underline{50.5}  & \underline{18.2}    & 13.3    & 15.4    \\ \hline
\bottomrule
\end{tabular}}
\end{table}

Table~\ref{tab:main_table} presents the comparative performance of unified embedding models and tool-based agentic systems across standard top-$k$ ranking metrics. We observe three key trends:
\begin{itemize}[leftmargin=10pt]
    \item \textbf{Visual-Linguistic Compression Loss.} 
    Caption-based text embedding pipelines consistently underperform multimodal embedding models. 
    Despite the augmentation of caption with structured metadata (\eg, time and location), the performance gap persists. This suggests a fundamental \textit{information bottleneck}: converting dense, fine-grained visual signals into discrete textual intermediates would lead to irreversible semantic loss, limiting the performance of text-based retrieval.
    
    \item \textbf{Superiority of Explicit Tool Orchestration.} 
    Tool-based agentic systems significantly outperform unified embedding models. This validates that personal album retrieval is not merely a visual matching problem but a \textit{multi-source constrained} problem. By explicitly orchestrating specialized tools, agents effectively bypass the limitations of monolithic embedding spaces, leveraging heterogeneous signals to resolve complex user intents.
    
    \item \textbf{Scaling Laws Hold for Both Paradigms.} 
    For embedding models, larger backbones yield better semantic alignment. Crucially, for agentic systems, performance gains correlate with both the backbone size and tool-calling capability (\eg, Qwen3-235B vs. Qwen3-8B). This indicates that retrieval quality in this regime is limited not just by visual perception, but by the \textit{planning capacity} required to decompose and execute multi-step queries.
    
\end{itemize}
Table~\ref{tab:phone_metrics} benchmarks the agentic retrieval systems against commercial mobile gallery systems using set-oriented metrics, separately on normal and zero-GT queries. 
This comparison reveals a critical trade-off between reasoning capability and system reliability:
\begin{itemize}[leftmargin=10pt]
    \item \textbf{Retrieval Ceiling (Normal Query).} 
    Agentic retrievers consistently achieve higher set-level F1 scores compared to all evaluated mobile systems on normal queries. 
    Potentially due to the on-device resource constraints, commercial retrieval engines usually prefer lightweight retrieval methods and thereby struggle with the entangled, intent-driven queries characteristic of \textsc{PhotoBench}. 
    Although costly, agents define a significantly higher performance ceiling for complex retrieval with their ability to synthesize evidence from multiple sources.

    \item \textbf{Reliability \& Abstention (Zero-GT Query):} 
    The trend reverses on rejection metrics. Mobile systems demonstrate superior \texttt{Reject-Recall}, reflecting a conservative engineering design optimized for precision (preferring no result over a wrong one). 
    In contrast, agentic systems exhibit a tendency towards \textit{Retrieval Hallucination}, forcing matches for non-existent queries. 
    This highlights a pivotal challenge for future research: \textbf{beyond maximizing recall, agentic retrievers must develop calibrated \textit{proactive abstention} mechanisms to operate reliably in open-world environments}.
\end{itemize}



\subsection{In-Depth Analysis}
\label{sec:in_depth_analysis}

To decompose the performance dynamics of each retrieval paradigm, we analyze the retrieval behaviors through the lens of our \textit{Source-Aware Query Taxonomy} introduced in Section~\ref{sec:query_taxonomy}. 
We derive the following key insights for the three retrieval paradigms.

\begin{table}[t]
\centering
\caption{
Recall@10 performance decomposition by source-aware query types. 
$\Delta$(A-M) and $\Delta$(A-T) denote performance gaps between agents and multimodal and caption-based text embedding models, respectively.
}
\label{tab:modality_gap}
\resizebox{0.47\textwidth}{!}{
\renewcommand\arraystretch{1.1}

\begin{tabular}{c|ccc|cc}
\toprule
\hline
Type & Agent(A) & Multimodal(M) & Text(T) & $\Delta$(A-M) & $\Delta$(A-T) \\
\hline
$S_V$ & 71.1 & 72.3 & 75.0 & -1.2 & -3.8 \\
$S_M$ & 57.9 & 7.2 & 7.9 & +50.8 & +50.1 \\
$S_F$ & 75.1 & 11.7 & 8.4 & +63.4 & +66.7 \\
$S_{VM}$ & 63.4 & 63.4 & 65.5 & +0.0 & -2.1 \\
$S_{VF}$ & 40.1 & 52.6 & 51.0 & -12.4 & -10.9 \\
$S_{MF}$ & 60.2 & 12.0 & 11.2 & +48.2 & +49.0 \\
$S_{VMF}$ & 37.5 & 51.4 & 46.5 & -13.9 & -9.0 \\
\hline
\bottomrule

\end{tabular}}
\end{table}

\subsubsection{Analysis of Unified Embedding Models}
\label{sec:modality gap analysis}

To diagnose the intrinsic limitations of unified embedding models, we analyze the averaged performance of three categories (\ie, agentic systems, multimodal embedding, and caption-based text embedding) across source-based query types defined in our taxonomy. 
We report Recall@10 performance in Table~\ref{tab:modality_gap} and derive two critical findings:
\begin{itemize}[leftmargin=10pt]
    \item \textbf{Modality Gap.}
    Unified embedding models exhibit a fundamental bias toward visual signals. While they achieve strong performance on purely visual queries ($S_V$), their efficacy collapses on queries requiring explicit metadata ($S_M$) or identity verification ($S_F$). The agentic system outperforms embedding models by vast margins in these categories. This confirms that current multimodal embeddings function primarily as \textit{visual similarity calculators}, lacking the capacity to encode precise spatio-temporal or social identity constraints in their latent space.

    \item \textbf{Visual-Anchor Effect.}
    Counterintuitively, embedding models remain competitive (often superior) on compositional queries containing visual terms (\ie, $S_{VM}$, $S_{VF}$, $S_{VMF}$), despite their demonstrated inability to process the non-visual components ($M$ and $F$). We attribute this to the \textit{visual-anchor effect}: in many compositional queries, the non-visual constraint is highly correlated with a distinctive visual cue (\eg, a ``birthday'' event implies the visual presence of a ``cake''). Embedding models exploit these visual anchors to retrieve correct targets via appearance matching without truly resolving the underlying metadata or identity logic. 
    In contrast, agentic systems act as strict logical reasoners and may fail if a specific tool misses a target, leading to lower recall on these ``visually solvable'' compositional queries.
\end{itemize}

\begin{table}[t]
\centering
\caption{
Tool ablation study on the best agentic system with Qwen3-235B-A22B. 
F1 scores are reported across query types as different types of tools are incrementally enabled. 
Note that the set composition tool ($T_S$) is always active.
}
\label{tab:tool ablation}
\vspace{-8pt}
\resizebox{\columnwidth}{!}{
\renewcommand\arraystretch{1.0}
\begin{tabular}{c|ccc|ccc|c}
\toprule
\hline
  Tools  & $S_V$    & $S_M$    & $S_F$    & $S_{VM}$   & $S_{VF}$   & $S_{MF}$   & $S_{VMF}$  \\ \hline
$T_V$   & 30.5 & 2.6  & 8.7  & 31.4 & 32.1 & 6.1  & 35.1 \\
$T_V+T_M$  & 32.3 & 54.7 & 12.9 & 34.1 & 33.2 & 25.5 & 33.4 \\
$T_V+T_F$ & 32.2 & 3.3  & 69.0 & 29.7 & 26.8 & 30.1 & 25.4 \\
$T_V + T_M+T_F$ & 36.9 & 69.7 & 90.1 & 44.7 & 30.1 & 76.8 & 32.5 \\ \hline
\bottomrule
\end{tabular}}
\end{table}

\subsubsection{Analysis of Agentic Retrieval Systems}

We conduct a tool ablation study to investigate the mechanism behind the agent's performance profile, specifically its strength on metadata/face queries versus its struggle on composite visual queries.
Using the strongest agent backbone (Qwen3-235B-A22B), we systematically enable different combinations of the Vector Search ($T_V$), Metadata Filter ($T_M$), and Face Engine ($T_F$) tools. 
Table~\ref{tab:tool ablation} reports the set-based F1 score across query types, from which we draw two key insights:
\begin{itemize}[leftmargin=10pt]
    \item \textbf{Decisive Role of Explicit Tool Access.}
    Performance improvements are strictly source-aligned, confirming that the agent's capability is architectural rather than emergent. 
    Enabling the Metadata Filter ($T_M$) yields a massive gain on $S_M$ queries (from 2.6 to 54.7), just as the Face Engine ($T_F$) unlocks $S_F$ performance (from 8.7 to 69.0). 
    Furthermore, visual grounding remains essential for disambiguation: introducing $T_V$ alongside non-visual tools is critical for composite queries like $S_{VMF}$, ensuring that the agent can visually pinpoint the correct target among dense near-duplicates that satisfy the metadata constraints.

    \item \textbf{Source Fusion Paradox}.
    Crucially, simply maximizing tool availability does not guarantee performance improvement. 
    For the most complex queries ($S_{VMF}$), enabling the full tool suite ($T_V + T_M + T_F$) yields a lower F1 score (32.5) than using the visual tool alone ($T_V$ at 35.1). 
    This counterintuitive degradation reveals the \textit{Source Fusion Paradox}: as the decision space expands, the agent increasingly struggles with \textit{tool orchestration}. 
    It often generates suboptimal execution plans or applies overly aggressive set intersections (\eg, combining a noisy face retrieval set with a precise time window), leading to the erroneous pruning of valid results.
    In this sense, the intrinsic reasoning and tool-calling capabilities are the major bottleneck of agentic retrieval systems.
\end{itemize}


\begin{table}[t]
\centering
\caption{
We report F1 scores of the best agent (Qwen3-235B-A22B) and mobile gallery systems as query complexity increases from single- to dual- and triple-source queries. 
$\Delta$ values indicate the performance degradation, exposing the Source Fusion Paradox in commercial engines.
}
\label{tab:fusion_paradox}
\vspace{-8pt}
\resizebox{0.47\textwidth}{!}{
\renewcommand\arraystretch{1.1}
\begin{tabular}{c|ccccccc}
\toprule
\hline
Type & Agent (Best) & Pho.A & Pho.B & Pho.C & Pho.D & Pho.E & Pho.F \\ \hline
$S_V$ & 37.6 & 33.2 & 20.5 & 34.6 & 34.1 & 40.6 & 36.4 \\
$S_M$ & 72.3 & 39.8 & 17.9 & 67.5 & 52.6 & 75.1 & 76.7 \\
$S_F$ & 93.4 & 1.8 & 1.1 & 1.7 & 6.7 & 0.0 & 10.3 \\
Avg. & 63.5 & 28.8 & 15.0 & 40.4 & 31.7 & 45.2 & 46.3 \\ \hline
$S_{VM}$ & 45.7 & 20.0 & 10.9 & 31.7 & 32.2 & 38.3 & 40.4 \\
$S_{VF}$ & 30.9 & 15.1 & 7.2 & 22.7 & 25.9 & 20.8 & 22.2 \\
$S_{MF}$ & 80.5 & 13.4 & 3.0 & 7.3 & 18.8 & 4.8 & 34.5 \\
Avg. & 52.4 & 16.2 & 7.1 & 20.6 & 25.6 & 21.3 & 32.4 \\ \hline
$S_{VMF}$ & 33.9 & 13.5 & 9.1 & 23.8 & 24.5 & 27.7 & 22.4 \\ \hline
$\Delta_{2-1}$ & -17.6 & -43.9 & -52.7 & -49.1 & -19.0 & -52.9 & -30.1 \\
$\Delta_{3-2}$  & -35.2 & -16.3 & +28.8 & +15.7 & -4.4 & +30.1 & -30.8 \\
$\Delta_{3-1}$ & -46.6 & -53.1 & -39.1 & -41.1 & -22.6 & -38.7 & -51.6 \\ \hline
\bottomrule
\end{tabular}}
\vspace{1mm}
\end{table}

\subsubsection{Analysis of Mobile Gallery Systems}
We analyze the performance degradation of six commercial gallery systems (Phones A--F) alongside our best agentic model. Table~\ref{tab:fusion_paradox} reports the average F1 scores across single, dual, and triple-source queries, along with the performance change ($\Delta$) between complexity levels. We highlight two divergent system behaviors:
\begin{itemize}[leftmargin=10pt]
    \item \textbf{Universal Degradation via Constraint Fusion ($\Delta_{2-1} < 0$).}
    The challenge of fusing heterogeneous sources is universal. When transitioning from single-source to dual-source queries (\eg, adding a time constraint to a visual search), both agentic and commercial systems suffer significant performance decay. 
    This confirms that the \textit{Source Fusion Paradox} is not an artifact of our agentic implementation, but a fundamental reliability gap in current retrieval architectures when handling conjoint constraints.

    \item \textbf{Rebound via Visual-Anchor Effect ($\Delta_{3-2} > 0$).}
    A distinct anomaly occurs at the triple-source level ($S_{VMF}$). While the agent continues to degrade monotonically, several commercial systems (Phones B, C, and E) exhibit a performance rebound, \textit{improving} by +15\% to +30\% compared to dual-source queries. 
    We attribute this to the same \textit{Visual Anchor Effect} observed in Section~\ref{sec:modality gap analysis}. 
    These commercial engines likely prioritize visual similarity scores over strict metadata or identity filtering. 
    When a visual term is re-introduced in a $S_{VMF}$ query, the system latches onto this visual anchor, effectively bypassing the failed non-visual logic ($S_{MF}$) to salvage recall. 
    Thus, this performance rebound is deceptive; it reflects a heuristic retreat to visual perception rather than a successful fusion of multi-source constraints.
\end{itemize}

\section{Conclusion and Future Direction}
\label{sec:conclusion}

We introduce \benchmark{}, a diagnostic benchmark that shifts the evaluation of mobile photo retrieval from visual matching to multi-source, intent-driven reasoning. 
By reconstructing the dense entanglement of visual semantics, spatial-temporal metadata, social identity, and temporal events established in authentic albums, \benchmark{} exposes critical limitations of existing retrieval models and systems that remain hidden in context-isolated web datasets.
Our investigation reveals two defining challenges for the research community like the modality gap and source fusion paradox. 
Ultimately, \benchmark{} suggests that the future of personal multimodal retrieval, especially for photo album scenarios, lies beyond establishing stronger unified embedding models. 
It requires a fundamental transition towards robust agentic reasoning systems capable of precise constraint satisfaction, proactive abstention, and the reliable fusion of heterogeneous, personalized signals.


\clearpage
\bibliographystyle{ACM-Reference-Format}
\bibliography{ref}

@inproceedings{lin2014microsoft,
  title={Microsoft coco: Common objects in context},
  author={Lin, Tsung-Yi and Maire, Michael and Belongie, Serge and Hays, James and Perona, Pietro and Ramanan, Deva and Doll{\'a}r, Piotr and Zitnick, C Lawrence},
  booktitle={European conference on computer vision},
  pages={740--755},
  year={2014},
  organization={Springer}
}

@inproceedings{plummer2015flickr30k,
  title={Flickr30k entities: Collecting region-to-phrase correspondences for richer image-to-sentence models},
  author={Plummer, Bryan A and Wang, Liwei and Cervantes, Chris M and Caicedo, Juan C and Hockenmaier, Julia and Lazebnik, Svetlana},
  booktitle={Proceedings of the IEEE international conference on computer vision},
  pages={2641--2649},
  year={2015}
}

@inproceedings{thrush2022winoground,
  title={Winoground: Probing vision and language models for visio-linguistic compositionality},
  author={Thrush, Tristan and Jiang, Ryan and Bartolo, Max and Singh, Amanpreet and Williams, Adina and Kiela, Douwe and Ross, Candace},
  booktitle={Proceedings of the IEEE/CVF Conference on Computer Vision and Pattern Recognition},
  pages={5238--5248},
  year={2022}
}

@inproceedings{liu2021visual,
  title={Visual news: Benchmark and challenges in news image captioning},
  author={Liu, Fuxiao and Wang, Yinghan and Wang, Tianlu and Ordonez, Vicente},
  booktitle={Proceedings of the 2021 conference on empirical methods in natural language processing},
  pages={6761--6771},
  year={2021}
}

@inproceedings{das2017visual,
  title={Visual dialog},
  author={Das, Abhishek and Kottur, Satwik and Gupta, Khushi and Singh, Avi and Yadav, Deshraj and Moura, Jos{\'e} MF and Parikh, Devi and Batra, Dhruv},
  booktitle={Proceedings of the IEEE conference on computer vision and pattern recognition},
  pages={326--335},
  year={2017}
}

@article{meng2025vlm2vec,
  title={Vlm2vec-v2: Advancing multimodal embedding for videos, images, and visual documents},
  author={Meng, Rui and Jiang, Ziyan and Liu, Ye and Su, Mingyi and Yang, Xinyi and Fu, Yuepeng and Qin, Can and Chen, Zeyuan and Xu, Ran and Xiong, Caiming and others},
  journal={arXiv preprint arXiv:2507.04590},
  year={2025}
}

@article{jiang2024vlm2vec,
  title={Vlm2vec: Training vision-language models for massive multimodal embedding tasks},
  author={Jiang, Ziyan and Meng, Rui and Yang, Xinyi and Yavuz, Semih and Zhou, Yingbo and Chen, Wenhu},
  journal={arXiv preprint arXiv:2410.05160},
  year={2024}
}

@article{vendrow2024inquire,
  title={INQUIRE: A natural world text-to-image retrieval benchmark},
  author={Vendrow, Edward and Pantazis, Omiros and Shepard, Alexander and Brostow, Gabriel and Jones, Kate E and Mac Aodha, Oisin and Beery, Sara and Van Horn, Grant},
  journal={Advances in Neural Information Processing Systems},
  volume={37},
  pages={126500--126514},
  year={2024}
}

@article{tschannen2025siglip,
  title={Siglip 2: Multilingual vision-language encoders with improved semantic understanding, localization, and dense features},
  author={Tschannen, Michael and Gritsenko, Alexey and Wang, Xiao and Naeem, Muhammad Ferjad and Alabdulmohsin, Ibrahim and Parthasarathy, Nikhil and Evans, Talfan and Beyer, Lucas and Xia, Ye and Mustafa, Basil and others},
  journal={arXiv preprint arXiv:2502.14786},
  year={2025}
}

@article{chen2024bge,
  title={Bge m3-embedding: Multi-lingual, multi-functionality, multi-granularity text embeddings through self-knowledge distillation},
  author={Chen, Jianlv and Xiao, Shitao and Zhang, Peitian and Luo, Kun and Lian, Defu and Liu, Zheng},
  journal={arXiv preprint arXiv:2402.03216},
  volume={4},
  number={5},
  year={2024}
}

@article{wang2024multilingual,
  title={Multilingual e5 text embeddings: A technical report},
  author={Wang, Liang and Yang, Nan and Huang, Xiaolong and Yang, Linjun and Majumder, Rangan and Wei, Furu},
  journal={arXiv preprint arXiv:2402.05672},
  year={2024}
}

@article{yang2025qwen3,
  title={Qwen3 technical report},
  author={Yang, An and Li, Anfeng and Yang, Baosong and Zhang, Beichen and Hui, Binyuan and Zheng, Bo and Yu, Bowen and Gao, Chang and Huang, Chengen and Lv, Chenxu and others},
  journal={arXiv preprint arXiv:2505.09388},
  year={2025}
}

@article{li2026qwen3,
  title={Qwen3-VL-Embedding and Qwen3-VL-Reranker: A Unified Framework for State-of-the-Art Multimodal Retrieval and Ranking},
  author={Li, Mingxin and Zhang, Yanzhao and Long, Dingkun and Chen, Keqin and Song, Sibo and Bai, Shuai and Yang, Zhibo and Xie, Pengjun and Yang, An and Liu, Dayiheng and others},
  journal={arXiv preprint arXiv:2601.04720},
  year={2026}
}

@misc{thirukovalluru2025breakingbatchbarrierb3,
      title={Breaking the Batch Barrier (B3) of Contrastive Learning via Smart Batch Mining}, 
      author={Raghuveer Thirukovalluru and Rui Meng and Ye Liu and Karthikeyan K and Mingyi Su and Ping Nie and Semih Yavuz and Yingbo Zhou and Wenhu Chen and Bhuwan Dhingra},
      year={2025},
      eprint={2505.11293},
      archivePrefix={arXiv},
      primaryClass={cs.CV},
      url={https://arxiv.org/abs/2505.11293}, 
}

@misc{OpenAI-o3,
      title={OpenAI o3 and o4-mini System Card}, 
      author={OpenAI},
      year={2025},
      url={https://cdn.openai.com/pdf/2221c875-02dc-4789-800b-e7758f3722c1/o3-and-o4-mini-system-card.pdf}, 
}

@misc{claude-sonnet,
      title={Claude Sonnet 4.5}, 
      author={Claude},
      year={2025},
      url={https://www.anthropic.com/claude/sonnet}, 
}

@misc{claude-opus,
      title={Claude Opus 4.5}, 
      author={Claude},
      year={2025},
      url={https://www-cdn.anthropic.com/bf10f64990cfda0ba858290be7b8cc6317685f47.pdf}, 
}

@misc{deepseekai2025deepseekv3technicalreport,
      title={DeepSeek-V3 Technical Report}, 
      author={DeepSeek-AI and Aixin Liu and Bei Feng and Bing Xue and Bingxuan Wang and Bochao Wu and Chengda Lu and Chenggang Zhao and Chengqi Deng and Chenyu Zhang and Chong Ruan and Damai Dai and Daya Guo and Dejian Yang and Deli Chen and Dongjie Ji and Erhang Li and Fangyun Lin and Fucong Dai and Fuli Luo and Guangbo Hao and Guanting Chen and Guowei Li and H. Zhang and Han Bao and Hanwei Xu and Haocheng Wang and Haowei Zhang and Honghui Ding and Huajian Xin and Huazuo Gao and Hui Li and Hui Qu and J. L. Cai and Jian Liang and Jianzhong Guo and Jiaqi Ni and Jiashi Li and Jiawei Wang and Jin Chen and Jingchang Chen and Jingyang Yuan and Junjie Qiu and Junlong Li and Junxiao Song and Kai Dong and Kai Hu and Kaige Gao and Kang Guan and Kexin Huang and Kuai Yu and Lean Wang and Lecong Zhang and Lei Xu and Leyi Xia and Liang Zhao and Litong Wang and Liyue Zhang and Meng Li and Miaojun Wang and Mingchuan Zhang and Minghua Zhang and Minghui Tang and Mingming Li and Ning Tian and Panpan Huang and Peiyi Wang and Peng Zhang and Qiancheng Wang and Qihao Zhu and Qinyu Chen and Qiushi Du and R. J. Chen and R. L. Jin and Ruiqi Ge and Ruisong Zhang and Ruizhe Pan and Runji Wang and Runxin Xu and Ruoyu Zhang and Ruyi Chen and S. S. Li and Shanghao Lu and Shangyan Zhou and Shanhuang Chen and Shaoqing Wu and Shengfeng Ye and Shengfeng Ye and Shirong Ma and Shiyu Wang and Shuang Zhou and Shuiping Yu and Shunfeng Zhou and Shuting Pan and T. Wang and Tao Yun and Tian Pei and Tianyu Sun and W. L. Xiao and Wangding Zeng and Wanjia Zhao and Wei An and Wen Liu and Wenfeng Liang and Wenjun Gao and Wenqin Yu and Wentao Zhang and X. Q. Li and Xiangyue Jin and Xianzu Wang and Xiao Bi and Xiaodong Liu and Xiaohan Wang and Xiaojin Shen and Xiaokang Chen and Xiaokang Zhang and Xiaosha Chen and Xiaotao Nie and Xiaowen Sun and Xiaoxiang Wang and Xin Cheng and Xin Liu and Xin Xie and Xingchao Liu and Xingkai Yu and Xinnan Song and Xinxia Shan and Xinyi Zhou and Xinyu Yang and Xinyuan Li and Xuecheng Su and Xuheng Lin and Y. K. Li and Y. Q. Wang and Y. X. Wei and Y. X. Zhu and Yang Zhang and Yanhong Xu and Yanhong Xu and Yanping Huang and Yao Li and Yao Zhao and Yaofeng Sun and Yaohui Li and Yaohui Wang and Yi Yu and Yi Zheng and Yichao Zhang and Yifan Shi and Yiliang Xiong and Ying He and Ying Tang and Yishi Piao and Yisong Wang and Yixuan Tan and Yiyang Ma and Yiyuan Liu and Yongqiang Guo and Yu Wu and Yuan Ou and Yuchen Zhu and Yuduan Wang and Yue Gong and Yuheng Zou and Yujia He and Yukun Zha and Yunfan Xiong and Yunxian Ma and Yuting Yan and Yuxiang Luo and Yuxiang You and Yuxuan Liu and Yuyang Zhou and Z. F. Wu and Z. Z. Ren and Zehui Ren and Zhangli Sha and Zhe Fu and Zhean Xu and Zhen Huang and Zhen Zhang and Zhenda Xie and Zhengyan Zhang and Zhewen Hao and Zhibin Gou and Zhicheng Ma and Zhigang Yan and Zhihong Shao and Zhipeng Xu and Zhiyu Wu and Zhongyu Zhang and Zhuoshu Li and Zihui Gu and Zijia Zhu and Zijun Liu and Zilin Li and Ziwei Xie and Ziyang Song and Ziyi Gao and Zizheng Pan},
      year={2025},
      eprint={2412.19437},
      archivePrefix={arXiv},
      primaryClass={cs.CL},
      url={https://arxiv.org/abs/2412.19437}, 
}

@article{shinn2023reflexion,
  title={Reflexion: Language agents with verbal reinforcement learning},
  author={Shinn, Noah and Cassano, Federico and Gopinath, Ashwin and Narasimhan, Karthik and Yao, Shunyu},
  journal={Advances in Neural Information Processing Systems},
  volume={36},
  pages={8634--8652},
  year={2023}
}

@misc{zhang2025qwen3embeddingadvancingtext,
      title={Qwen3 Embedding: Advancing Text Embedding and Reranking Through Foundation Models}, 
      author={Yanzhao Zhang and Mingxin Li and Dingkun Long and Xin Zhang and Huan Lin and Baosong Yang and Pengjun Xie and An Yang and Dayiheng Liu and Junyang Lin and Fei Huang and Jingren Zhou},
      year={2025},
      eprint={2506.05176},
      archivePrefix={arXiv},
      primaryClass={cs.CL},
      url={https://arxiv.org/abs/2506.05176}, 
}

@misc{liang2025mmr1unleashingpowerunified,
      title={MM-R1: Unleashing the Power of Unified Multimodal Large Language Models for Personalized Image Generation}, 
      author={Qian Liang and Yujia Wu and Kuncheng Li and Jiwei Wei and Shiyuan He and Jinyu Guo and Ning Xie},
      year={2025},
      eprint={2508.11433},
      archivePrefix={arXiv},
      primaryClass={cs.CV},
      url={https://arxiv.org/abs/2508.11433}, 
}

@misc{wu2025mmsearchr1incentivizinglmmssearch,
      title={MMSearch-R1: Incentivizing LMMs to Search}, 
      author={Jinming Wu and Zihao Deng and Wei Li and Yiding Liu and Bo You and Bo Li and Zejun Ma and Ziwei Liu},
      year={2025},
      eprint={2506.20670},
      archivePrefix={arXiv},
      primaryClass={cs.CV},
      url={https://arxiv.org/abs/2506.20670}, 
}

@misc{jiang2024mmsearchbenchmarkingpotentiallarge,
      title={MMSearch: Benchmarking the Potential of Large Models as Multi-modal Search Engines}, 
      author={Dongzhi Jiang and Renrui Zhang and Ziyu Guo and Yanmin Wu and Jiayi Lei and Pengshuo Qiu and Pan Lu and Zehui Chen and Chaoyou Fu and Guanglu Song and Peng Gao and Yu Liu and Chunyuan Li and Hongsheng Li},
      year={2024},
      eprint={2409.12959},
      archivePrefix={arXiv},
      primaryClass={cs.CV},
      url={https://arxiv.org/abs/2409.12959}, 
}

@misc{Ops-MM,
      title={Ops-MM-embedding-v1-7B}, 
      author={OpenSearch-AI},
      year={2025},
      url={https://huggingface.co/OpenSearch-AI/Ops-MM-embedding-v1-7B}, 
}

@misc{openai2024gpt4ocard,
      title={GPT-4o System Card}, 
      author={OpenAI and : and Aaron Hurst and Adam Lerer and Adam P. Goucher and Adam Perelman and Aditya Ramesh and Aidan Clark and AJ Ostrow and Akila Welihinda and Alan Hayes and Alec Radford and Aleksander Mądry and Alex Baker-Whitcomb and Alex Beutel and Alex Borzunov and Alex Carney and Alex Chow and Alex Kirillov and Alex Nichol and Alex Paino and Alex Renzin and Alex Tachard Passos and Alexander Kirillov and Alexi Christakis and Alexis Conneau and Ali Kamali and Allan Jabri and Allison Moyer and Allison Tam and Amadou Crookes and Amin Tootoochian and Amin Tootoonchian and Ananya Kumar and Andrea Vallone and Andrej Karpathy and Andrew Braunstein and Andrew Cann and Andrew Codispoti and Andrew Galu and Andrew Kondrich and Andrew Tulloch and Andrey Mishchenko and Angela Baek and Angela Jiang and Antoine Pelisse and Antonia Woodford and Anuj Gosalia and Arka Dhar and Ashley Pantuliano and Avi Nayak and Avital Oliver and Barret Zoph and Behrooz Ghorbani and Ben Leimberger and Ben Rossen and Ben Sokolowsky and Ben Wang and Benjamin Zweig and Beth Hoover and Blake Samic and Bob McGrew and Bobby Spero and Bogo Giertler and Bowen Cheng and Brad Lightcap and Brandon Walkin and Brendan Quinn and Brian Guarraci and Brian Hsu and Bright Kellogg and Brydon Eastman and Camillo Lugaresi and Carroll Wainwright and Cary Bassin and Cary Hudson and Casey Chu and Chad Nelson and Chak Li and Chan Jun Shern and Channing Conger and Charlotte Barette and Chelsea Voss and Chen Ding and Cheng Lu and Chong Zhang and Chris Beaumont and Chris Hallacy and Chris Koch and Christian Gibson and Christina Kim and Christine Choi and Christine McLeavey and Christopher Hesse and Claudia Fischer and Clemens Winter and Coley Czarnecki and Colin Jarvis and Colin Wei and Constantin Koumouzelis and Dane Sherburn and Daniel Kappler and Daniel Levin and Daniel Levy and David Carr and David Farhi and David Mely and David Robinson and David Sasaki and Denny Jin and Dev Valladares and Dimitris Tsipras and Doug Li and Duc Phong Nguyen and Duncan Findlay and Edede Oiwoh and Edmund Wong and Ehsan Asdar and Elizabeth Proehl and Elizabeth Yang and Eric Antonow and Eric Kramer and Eric Peterson and Eric Sigler and Eric Wallace and Eugene Brevdo and Evan Mays and Farzad Khorasani and Felipe Petroski Such and Filippo Raso and Francis Zhang and Fred von Lohmann and Freddie Sulit and Gabriel Goh and Gene Oden and Geoff Salmon and Giulio Starace and Greg Brockman and Hadi Salman and Haiming Bao and Haitang Hu and Hannah Wong and Haoyu Wang and Heather Schmidt and Heather Whitney and Heewoo Jun and Hendrik Kirchner and Henrique Ponde de Oliveira Pinto and Hongyu Ren and Huiwen Chang and Hyung Won Chung and Ian Kivlichan and Ian O'Connell and Ian O'Connell and Ian Osband and Ian Silber and Ian Sohl and Ibrahim Okuyucu and Ikai Lan and Ilya Kostrikov and Ilya Sutskever and Ingmar Kanitscheider and Ishaan Gulrajani and Jacob Coxon and Jacob Menick and Jakub Pachocki and James Aung and James Betker and James Crooks and James Lennon and Jamie Kiros and Jan Leike and Jane Park and Jason Kwon and Jason Phang and Jason Teplitz and Jason Wei and Jason Wolfe and Jay Chen and Jeff Harris and Jenia Varavva and Jessica Gan Lee and Jessica Shieh and Ji Lin and Jiahui Yu and Jiayi Weng and Jie Tang and Jieqi Yu and Joanne Jang and Joaquin Quinonero Candela and Joe Beutler and Joe Landers and Joel Parish and Johannes Heidecke and John Schulman and Jonathan Lachman and Jonathan McKay and Jonathan Uesato and Jonathan Ward and Jong Wook Kim and Joost Huizinga and Jordan Sitkin and Jos Kraaijeveld and Josh Gross and Josh Kaplan and Josh Snyder and Joshua Achiam and Joy Jiao and Joyce Lee and Juntang Zhuang and Justyn Harriman and Kai Fricke and Kai Hayashi and Karan Singhal and Katy Shi and Kavin Karthik and Kayla Wood and Kendra Rimbach and Kenny Hsu and Kenny Nguyen and Keren Gu-Lemberg and Kevin Button and Kevin Liu and Kiel Howe and Krithika Muthukumar and Kyle Luther and Lama Ahmad and Larry Kai and Lauren Itow and Lauren Workman and Leher Pathak and Leo Chen and Li Jing and Lia Guy and Liam Fedus and Liang Zhou and Lien Mamitsuka and Lilian Weng and Lindsay McCallum and Lindsey Held and Long Ouyang and Louis Feuvrier and Lu Zhang and Lukas Kondraciuk and Lukasz Kaiser and Luke Hewitt and Luke Metz and Lyric Doshi and Mada Aflak and Maddie Simens and Madelaine Boyd and Madeleine Thompson and Marat Dukhan and Mark Chen and Mark Gray and Mark Hudnall and Marvin Zhang and Marwan Aljubeh and Mateusz Litwin and Matthew Zeng and Max Johnson and Maya Shetty and Mayank Gupta and Meghan Shah and Mehmet Yatbaz and Meng Jia Yang and Mengchao Zhong and Mia Glaese and Mianna Chen and Michael Janner and Michael Lampe and Michael Petrov and Michael Wu and Michele Wang and Michelle Fradin and Michelle Pokrass and Miguel Castro and Miguel Oom Temudo de Castro and Mikhail Pavlov and Miles Brundage and Miles Wang and Minal Khan and Mira Murati and Mo Bavarian and Molly Lin and Murat Yesildal and Nacho Soto and Natalia Gimelshein and Natalie Cone and Natalie Staudacher and Natalie Summers and Natan LaFontaine and Neil Chowdhury and Nick Ryder and Nick Stathas and Nick Turley and Nik Tezak and Niko Felix and Nithanth Kudige and Nitish Keskar and Noah Deutsch and Noel Bundick and Nora Puckett and Ofir Nachum and Ola Okelola and Oleg Boiko and Oleg Murk and Oliver Jaffe and Olivia Watkins and Olivier Godement and Owen Campbell-Moore and Patrick Chao and Paul McMillan and Pavel Belov and Peng Su and Peter Bak and Peter Bakkum and Peter Deng and Peter Dolan and Peter Hoeschele and Peter Welinder and Phil Tillet and Philip Pronin and Philippe Tillet and Prafulla Dhariwal and Qiming Yuan and Rachel Dias and Rachel Lim and Rahul Arora and Rajan Troll and Randall Lin and Rapha Gontijo Lopes and Raul Puri and Reah Miyara and Reimar Leike and Renaud Gaubert and Reza Zamani and Ricky Wang and Rob Donnelly and Rob Honsby and Rocky Smith and Rohan Sahai and Rohit Ramchandani and Romain Huet and Rory Carmichael and Rowan Zellers and Roy Chen and Ruby Chen and Ruslan Nigmatullin and Ryan Cheu and Saachi Jain and Sam Altman and Sam Schoenholz and Sam Toizer and Samuel Miserendino and Sandhini Agarwal and Sara Culver and Scott Ethersmith and Scott Gray and Sean Grove and Sean Metzger and Shamez Hermani and Shantanu Jain and Shengjia Zhao and Sherwin Wu and Shino Jomoto and Shirong Wu and Shuaiqi and Xia and Sonia Phene and Spencer Papay and Srinivas Narayanan and Steve Coffey and Steve Lee and Stewart Hall and Suchir Balaji and Tal Broda and Tal Stramer and Tao Xu and Tarun Gogineni and Taya Christianson and Ted Sanders and Tejal Patwardhan and Thomas Cunninghman and Thomas Degry and Thomas Dimson and Thomas Raoux and Thomas Shadwell and Tianhao Zheng and Todd Underwood and Todor Markov and Toki Sherbakov and Tom Rubin and Tom Stasi and Tomer Kaftan and Tristan Heywood and Troy Peterson and Tyce Walters and Tyna Eloundou and Valerie Qi and Veit Moeller and Vinnie Monaco and Vishal Kuo and Vlad Fomenko and Wayne Chang and Weiyi Zheng and Wenda Zhou and Wesam Manassra and Will Sheu and Wojciech Zaremba and Yash Patil and Yilei Qian and Yongjik Kim and Youlong Cheng and Yu Zhang and Yuchen He and Yuchen Zhang and Yujia Jin and Yunxing Dai and Yury Malkov},
      year={2024},
      eprint={2410.21276},
      archivePrefix={arXiv},
      primaryClass={cs.CL},
      url={https://arxiv.org/abs/2410.21276}, 
}

@inproceedings{gurrin2023introduction,
  title={Introduction to the sixth annual lifelog search challenge, LSC’23},
  author={Gurrin, Cathal and J{\'o}nsson, Bj{\"o}rn {\TH}{\'o}r and Nguyen, Duc Tien Dang and Healy, Graham and Lokoc, Jakub and Zhou, Liting and Rossetto, Luca and Tran, Minh-Triet and H{\"u}rst, Wolfgang and Bailer, Werner and others},
  booktitle={Proceedings of the 2023 ACM International Conference on Multimedia Retrieval},
  pages={678--679},
  year={2023}
}

@inproceedings{yao2022react,
  title={React: Synergizing reasoning and acting in language models},
  author={Yao, Shunyu and Zhao, Jeffrey and Yu, Dian and Du, Nan and Shafran, Izhak and Narasimhan, Karthik R and Cao, Yuan},
  booktitle={The eleventh international conference on learning representations},
  year={2022}
}

@article{kirichenko2025abstentionbench,
  title={AbstentionBench: Reasoning LLMs Fail on Unanswerable Questions},
  author={Kirichenko, Polina and Ibrahim, Mark and Chaudhuri, Kamalika and Bell, Samuel J},
  journal={arXiv preprint arXiv:2506.09038},
  year={2025}
}

@inproceedings{
  liu2025toolace,
  title={Tool{ACE}: Winning the Points of {LLM} Function Calling},
  author={Weiwen Liu and Xu Huang and Xingshan Zeng and xinlong hao and Shuai Yu and Dexun Li and Shuai Wang and Weinan Gan and Zhengying Liu and Yuanqing Yu and Zezhong WANG and Yuxian Wang and Wu Ning and Yutai Hou and Bin Wang and Chuhan Wu and Wang Xinzhi and Yong Liu and Yasheng Wang and Duyu Tang and Dandan Tu and Lifeng Shang and Xin Jiang and Ruiming Tang and Defu Lian and Qun Liu and Enhong Chen},
  booktitle={The Thirteenth International Conference on Learning Representations},
  year={2025},
  url={https://openreview.net/forum?id=8EB8k6DdCU}
}

@inproceedings{radford2021learning,
  title={Learning transferable visual models from natural language supervision},
  author={Radford, Alec and Kim, Jong Wook and Hallacy, Chris and Ramesh, Aditya and Goh, Gabriel and Agarwal, Sandhini and Sastry, Girish and Askell, Amanda and Mishkin, Pamela and Clark, Jack and others},
  booktitle={International conference on machine learning},
  pages={8748--8763},
  year={2021},
  organization={PMLR}
}

@inproceedings{zhai2023sigmoid,
  title={Sigmoid loss for language image pre-training},
  author={Zhai, Xiaohua and Mustafa, Basil and Kolesnikov, Alexander and Beyer, Lucas},
  booktitle={Proceedings of the IEEE/CVF International Conference on Computer Vision},
  pages={11975--11986},
  year={2023}
}

@article{fashioniq,
  title={The Fashion IQ Dataset: Retrieving Images by Combining Side Information and Relative Natural Language Feedback},
  author={Wu, Hui and Gao, Yupeng and Guo, Xiaoxiao and Al-Halah, Ziad and Rennie, Steven and Grauman, Kristen and Feris, Rogerio},
  journal={CVPR},
  year={2021}
}

@InProceedings{cirr,
  author    = {Liu, Zheyuan and Rodriguez-Opazo, Cristian and Teney, Damien and Gould, Stephen},
  title     = {Image Retrieval on Real-Life Images With Pre-Trained Vision-and-Language Models},
  booktitle = {Proceedings of the IEEE/CVF International Conference on Computer Vision (ICCV)},
  year      = {2021},
  pages     = {2125-2134}
}

@inproceedings{autocir,
  title={Generative Thinking, Corrective Action: User-Friendly Composed Image Retrieval via Automatic Multi-Agent Collaboration},
  author={Cheng, Zhangtao and Ma, Yuhao and Lang, Jian and Zhang, Kunpeng and Zhong, Ting and Wang, Yong and Zhou, Fan},
  booktitle={Proceedings of the 31st ACM SIGKDD Conference on Knowledge Discovery and Data Mining (KDD)},
  pages={334--344},
  year={2025}
}

@article{xr,
  title={XR: Cross-Modal Agents for Composed Image Retrieval},
  author={Yang, Zhongyu and Pang, Wei and Yuan, Yingfang},
  journal={arXiv preprint arXiv:2601.14245},
  year={2026}
}

@inproceedings{ldre,
  title={LDRE: LLM-based Divergent Reasoning and Ensemble for Zero-Shot Composed Image Retrieval},
  author={Yang, Zhenyu and Xue, Dizhan and Qian, Shengsheng and Dong, Weiming and Xu, Changsheng},
  booktitle={Proceedings of the 47th International ACM SIGIR Conference on Research and Development in Information Retrieval (SIGIR)},
  pages={80--90},
  year={2024}
}

@article{MRACIR,
  title={Multimodal Reasoning Agent for Zero-Shot Composed Image Retrieval},
  author={Tu, Rong-Cheng and Sun, Wenhao and You, Hanzhe and Wang, Yingjie and Huang, Jiaxing and Shen, Li and Tao, Dacheng},
  journal={arXiv preprint arXiv:2505.19952},
  year={2025}
}

@inproceedings{Reason-before-retrieve,
  title={Reason-before-retrieve: One-stage reflective chain-of-thoughts for training-free zero-shot composed image retrieval},
  author={Tang, Yuanmin and Zhang, Jue and Qin, Xiaoting and Yu, Jing and Gou, Gaopeng and Xiong, Gang and Lin, Qingwei and Rajmohan, Saravan and Zhang, Dongmei and Wu, Qi},
  booktitle={Proceedings of the IEEE/CVF Conference on Computer Vision and Pattern Recognition (CVPR)},
  pages={14400--14410},
  year={2025}
}

@article{searle,
  title={iSEARLE: Improving Textual Inversion for Zero-Shot Composed Image Retrieval}, 
  author={Agnolucci, Lorenzo and Baldrati, Alberto and Bertini, Marco and Del Bimbo, Alberto},
  journal={arXiv preprint arXiv:2405.02951},
  year={2024}
}

@article{cireVL,
  title={Vision-by-Language for Training-Free Compositional Image Retrieval},
  author={Shyamgopal Karthik and Karsten Roth and Massimiliano Mancini and Zeynep Akata},
  journal={International Conference on Learning Representations (ICLR)},
  year={2024}
}

@inproceedings{lincir,
  title={Language-only Training of Zero-shot Composed Image Retrieval},
  author={Gu, Geonmo and Chun, Sanghyuk and Kim, Wonjae and Kang, Yoohoon and Yun, Sangdoo},
  booktitle={Proceedings of the IEEE/CVF Conference on Computer Vision and Pattern Recognition (CVPR)},
  year={2024}
}

@inproceedings{fire,
  title={FiRE: Enhancing MLLMs with fine-grained context learning for complex image retrieval},
  author={Hou, Bohan and Lin, Haoqiang and Song, Xuemeng and Wen, Haokun and Liu, Meng and Hu, Yupeng and Zhao, Xiangyu},
  booktitle={Proceedings of the 48th International ACM SIGIR Conference on Research and Development in Information Retrieval (SIGIR)},
  pages={803--812},
  year={2025}
}

@article{thirukovalluru2025breaking,
  title={Breaking the Batch Barrier (B3) of Contrastive Learning via Smart Batch Mining},
  author={Thirukovalluru, Raghuveer and Meng, Rui and Liu, Ye and Su, Mingyi and Nie, Ping and Yavuz, Semih and Zhou, Yingbo and Chen, Wenhu and Dhingra, Bhuwan and others},
  journal={arXiv preprint arXiv:2505.11293},
  year={2025}
}

@article{jian2025rzenembed,
  title={Rzenembed: Towards comprehensive multimodal retrieval},
  author={Jian, Weijian and Zhang, Yajun and Liang, Dawei and Xie, Chunyu and He, Yixiao and Leng, Dawei and Yin, Yuhui},
  journal={arXiv preprint arXiv:2510.27350},
  year={2025}
}

@inproceedings{chen2024internvl,
  title={Internvl: Scaling up vision foundation models and aligning for generic visual-linguistic tasks},
  author={Chen, Zhe and Wu, Jiannan and Wang, Wenhai and Su, Weijie and Chen, Guo and Xing, Sen and Zhong, Muyan and Zhang, Qinglong and Zhu, Xizhou and Lu, Lewei and others},
  booktitle={Proceedings of the IEEE/CVF conference on computer vision and pattern recognition},
  pages={24185--24198},
  year={2024}
}

@misc{xue2025improvemultimodalembeddinglearning,
      title={Improve Multi-Modal Embedding Learning via Explicit Hard Negative Gradient Amplifying}, 
      author={Youze Xue and Dian Li and Gang Liu},
      year={2025},
      eprint={2506.02020},
      archivePrefix={arXiv},
      primaryClass={cs.CV},
      url={https://arxiv.org/abs/2506.02020}, 
}

@article{lu2025internvl,
  title={Internvl-x: Advancing and accelerating internvl series with efficient visual token compression},
  author={Lu, Dongchen and Sun, Yuyao and Zhang, Zilu and Huang, Leping and Zeng, Jianliang and Shu, Mao and Cao, Huo},
  journal={arXiv preprint arXiv:2503.21307},
  year={2025}
}

@article{schuhmann2022laion,
  title={Laion-5b: An open large-scale dataset for training next generation image-text models},
  author={Schuhmann, Christoph and Beaumont, Romain and Vencu, Richard and Gordon, Cade and Wightman, Ross and Cherti, Mehdi and Coombes, Theo and Katta, Aarush and Mullis, Clayton and Wortsman, Mitchell and others},
  journal={Advances in neural information processing systems},
  volume={35},
  pages={25278--25294},
  year={2022}
}

@article{ma2024unifying,
  title={Unifying multimodal retrieval via document screenshot embedding},
  author={Ma, Xueguang and Lin, Sheng-Chieh and Li, Minghan and Chen, Wenhu and Lin, Jimmy},
  journal={arXiv preprint arXiv:2406.11251},
  year={2024}
}

\clearpage
\appendix
\section{Details of Dataset Construction}
Here we provide the complete prompts and detailed methods for each stage of the \benchmark{} construction pipeline. All prompts are designed both in Chinese and in English. Here we only show the English version
\label{app:prompts}

\tcbset{
    promptstyle/.style={
        colback=backorange,
        colframe=frameorange,
        fonttitle=\bfseries,
        arc=2pt,
        breakable, 
        left=2mm, right=2mm, top=2mm, bottom=2mm,
        boxrule=0.5pt,
        fontupper=\small,
        fontlower=\small
    }
}

\subsection{Event Narrative Update Prompt}
\begin{tcolorbox}[promptstyle, title=Incremental Event Narrative Synthesis]
\textbf{Role:} Precise Event Analyst.
\textbf{Objective:} Iteratively refine a coherent narrative from clustered images.

\textbf{Logic Protocol:}
\begin{itemize}[leftmargin=1.2em, nosep]
    \item \textbf{Initial (Index 0):} Establish baseline hypothesis from visual cues.
    \item \textbf{Update (Intermediate):} Synthesize new details or advance storyline using temporal connectors (e.g., ``then'', ``afterward''). Correct earlier assumptions if contradictions arise.
    \item \textbf{Final (Index $N-1$):} Produce a polished, globally consistent summary.
\end{itemize}

\textbf{Constraints:} 
Continuous narrative (no lists); Evidence-based inference (e.g., cake $\to$ birthday); Eliminate redundancy; Retroactively refine for causal coherence.
\end{tcolorbox}

\subsection{Intention Inference Prompt}
\begin{tcolorbox}[promptstyle, title=Context-Aware Capture Intention Inference]
\textbf{Objective:} Infer the primary motivation behind a photo by synthesizing behavioral trajectory and visual content.

\textbf{Analysis Framework:}
\begin{enumerate}[leftmargin=1.2em, nosep]
    \item \textbf{Trajectory Mapping:} Analyze the user's preceding activity sequence.
    \item \textbf{Role Identification:} Define the photo's functional role (milestone, routine, or climax) within the event chain.
    \item \textbf{Motivation Mapping:} Categorize intent (e.g., memory preservation, information retrieval, artistic expression, or social sharing).
\end{enumerate}

\textbf{Output:} A single declarative sentence (e.g., ``Recording a specific exhibit detail for future reference''). No meta-talk or labels.
\end{tcolorbox}

\subsection{Query Generation Prompt}
\begin{tcolorbox}[promptstyle, title=Intent-Driven Multimodal Query Generation]
\textbf{Task:} Generate 3 diverse, colloquial queries reflecting the user's search psychological state.

\textbf{Generation Dimensions:}
\begin{itemize}[leftmargin=1.2em, nosep]
    \item \textbf{Event-centric:} Focus on the macro-activity (e.g., ``beach camping'').
    \item \textbf{Object-centric:} Focus on key subjects and states (e.g., ``cat sunbathing'').
    \item \textbf{Freeform:} Focus on atmosphere, color, or specific metadata constraints.
\end{itemize}

\textbf{Requirements:} 
Incorporate person nicknames; Use metadata naturally; Keep queries $<15$ characters; Omit filler words like ``photo of''.

\textbf{Output Format:} JSON with \texttt{queries} and \texttt{reasoning} (mapping intent to query logic).
\end{tcolorbox}

\subsection{Zero-GT queries Synthesis Method}
\label{app:zero_gt_construct}
We generate zero-GT queries in two ways:
\begin{itemize}[leftmargin=10pt]
    \item \textbf{Metadata Perturbation and Semantic Variation:} We employ metadata perturbation by introducing contradictions such as conflicting timestamps, and modify key relational roles (\eg, changing family roles) to create plausible but non-existent descriptions. For semantic variation, we alter key elements in the query, such as modifying relationships or context, making the description realistic but still without a corresponding match in the dataset.
    
    \item \textbf{Query Variations:} We also create variations on existing queries by three methods:
        (1) Entity Mismatch: Replace the subject in a query with a rare or unrelated entity (\eg, changing \textit{birthday party with Alice} to \textit{birthday party with a celebrity}).
        (2) Scene Mismatch: Place individuals in unusual or unexpected settings, such as an unlikely career or an unfamiliar outdoor scene (\eg, \textit{changing meeting at a restaurant with colleagues} to \textit{meeting in a foreign office}).
        (3) Detail Enhancement: Add specific or uncommon details to a query, such as including rare actions or unique props, making the matching process significantly more difficult .
\end{itemize}
These zero-GT queries undergo the same verification process described in Section~\ref{sec:candidate_mining_verification}, where human experts confirm that no relevant ground-truth images exist in the dataset for these queries.

\subsection{Zero-GT Query Generation Prompt}
\begin{tcolorbox}[promptstyle, title=Zero-Ground Truth (Zero-GT) Synthesis]
\textbf{Goal:} Synthesize "unfulfillable" queries that are contextually plausible but lack visual matches to test system rejection capabilities.

\textbf{Methodology:}
\begin{itemize}[leftmargin=1.2em, nosep]
    \item \textbf{M1: Semantic Deviation:} Shift entities or scenes to realistic but absent variants; hyper-specify details to make matches statistically impossible.
    \item \textbf{M2: Constraint Injection:} Append conflicting metadata to valid queries (e.g., a year the user never visited that location, or a location where the specific activity never occurred).
\end{itemize}

\textbf{Constraints:} Maintain realism (no sci-fi); Colloquial style; Verified via exhaustive VL-model cross-check to ensure zero matches.
\end{tcolorbox}

\section{Implementation Details of Metrics}

\subsection{Metrics for Query Linguistic Analysis}
\label{appendix:lingui_metric}

To facilitate the reproduction of our linguistic analysis of queries, we provide the formal definitions and algorithmic implementations for the metrics presented in Figure~\ref{fig:dense_summary}. All metrics are computed using the \texttt{en\_core\_web\_sm} model from the \texttt{spaCy} NLP library.

\subsubsection{Average Query Length and Noun Density}
The \textbf{Average Query Length} is defined as the mean number of tokens per query. \textbf{Noun Density} ($\rho_{n}$) measures the concentration of informational entities within a query, calculated as the ratio of nouns and proper nouns to the total token count:
$$ \rho_{n} = \frac{\sum_{i \in T} \mathbb{I}(pos(i) \in \{\text{NOUN, PROPN}\})}{|T|} $$
where $T$ is the set of tokens in a query, and $\mathbb{I}(\cdot)$ is the indicator function.

\subsubsection{Average Syntactic Depth}
\textbf{Syntactic Depth} quantifies the structural complexity of a query. For each input, we generate a dependency parse tree. The depth of a tree $D$ is the maximum path length from the root node to any leaf node:
$$ D = \max_{n \in \text{nodes}} (\text{path\_length}(\text{root}, n)) $$
For multi-sentence queries, we record the maximum depth across all constituent sentences. A lower depth indicates a flatter, more fragmented grammatical structure typical of search-style interactions.

\subsubsection{Lexical Diversity (MTLD)}
We adopt the \textbf{Measure of Textual Lexical Diversity (MTLD)} to evaluate vocabulary richness. Unlike the simple Type-Token Ratio (TTR), MTLD is robust against variations in total corpus length. 

The calculation follows a sequential factor-based approach:
\begin{enumerate}[leftmargin=12pt]
    \item \textbf{Factorization:} The algorithm traverses the token sequence until the TTR drops below a predefined threshold (set to $0.72$ in our implementation). At this point, a ``factor'' is completed, and the TTR calculation resets.
    \item \textbf{Bidirectional Scoring:} This process is performed both forward ($F_{fwd}$) and backward ($F_{bwd}$) through the corpus to ensure stability.
    \item \textbf{Final Calculation:} The MTLD score is the ratio of the total number of tokens $N$ to the average number of factors:
    $$ \text{MTLD} = \frac{N}{(F_{fwd} + F_{bwd}) / 2} $$
\end{enumerate}
A higher MTLD score signifies that the dataset utilizes a broader and more specialized vocabulary.

\subsection{Metrics for System Rejection Ability Evaluation}
\label{appendix: rejection metrics}


To evaluate the rejection capability of a system, we regard an \emph{empty prediction} as a rejection event.
Given a test set $\{(g_i, p_i)\}_{i=1}^{N}$, let $\mathbb{I}(\cdot)$ be the indicator function and define
\[
E(x)=\mathbb{I}(x=\varnothing).
\]
We compute the following counts:
\[
A=\sum_{i=1}^{N}\mathbb{I}\big(E(g_i)=1 \land E(p_i)=1\big), \quad
B=\sum_{i=1}^{N}\mathbb{I}\big(E(g_i)=0 \land E(p_i)=1\big),
\]
\[
C=\sum_{i=1}^{N}\mathbb{I}\big(E(g_i)=1 \land E(p_i)=0\big), \quad
D=\sum_{i=1}^{N}\mathbb{I}\big(E(g_i)=0 \land E(p_i)=0\big),
\]
where $A$ is correct rejection, $B$ is false rejection, $C$ is missed rejection, and $D$ is correct non-rejection.

\textbf{Reject-Precision}, \textbf{Reject-Recall}, and \textbf{Reject-F1} are defined as:
\[
\mathrm{Reject\text{-}Precision}=\frac{A}{A+B}, \qquad
\mathrm{Reject\text{-}Recall}=\frac{A}{A+C},
\]
\[
\mathrm{Reject\text{-}F1}
=\frac{2\cdot \mathrm{Reject\text{-}Precision}\cdot \mathrm{Reject\text{-}Recall}}
{\mathrm{Reject\text{-}Precision}+\mathrm{Reject\text{-}Recall}}
=\frac{2A}{2A+B+C}.
\]

Reject-Precision measures the \emph{correctness} of rejection decisions (\ie, how often rejected cases truly have empty ground truth),
Reject-Recall measures the \emph{coverage} of rejection (\ie, how many empty-ground-truth cases are successfully rejected),
and Reject-F1 provides a single summary by \emph{balancing} Reject-Precision and Reject-Recall, penalizing both over-rejection ($B$) and under-rejection ($C$).

\section{Extended Dataset Statistics}
\label{app:stats}

\subsection{Comparsion with classic benchmarks}
\label{app:comp_benches}
\begin{figure}[t]
    \centering
    \includegraphics[width=0.98\columnwidth]{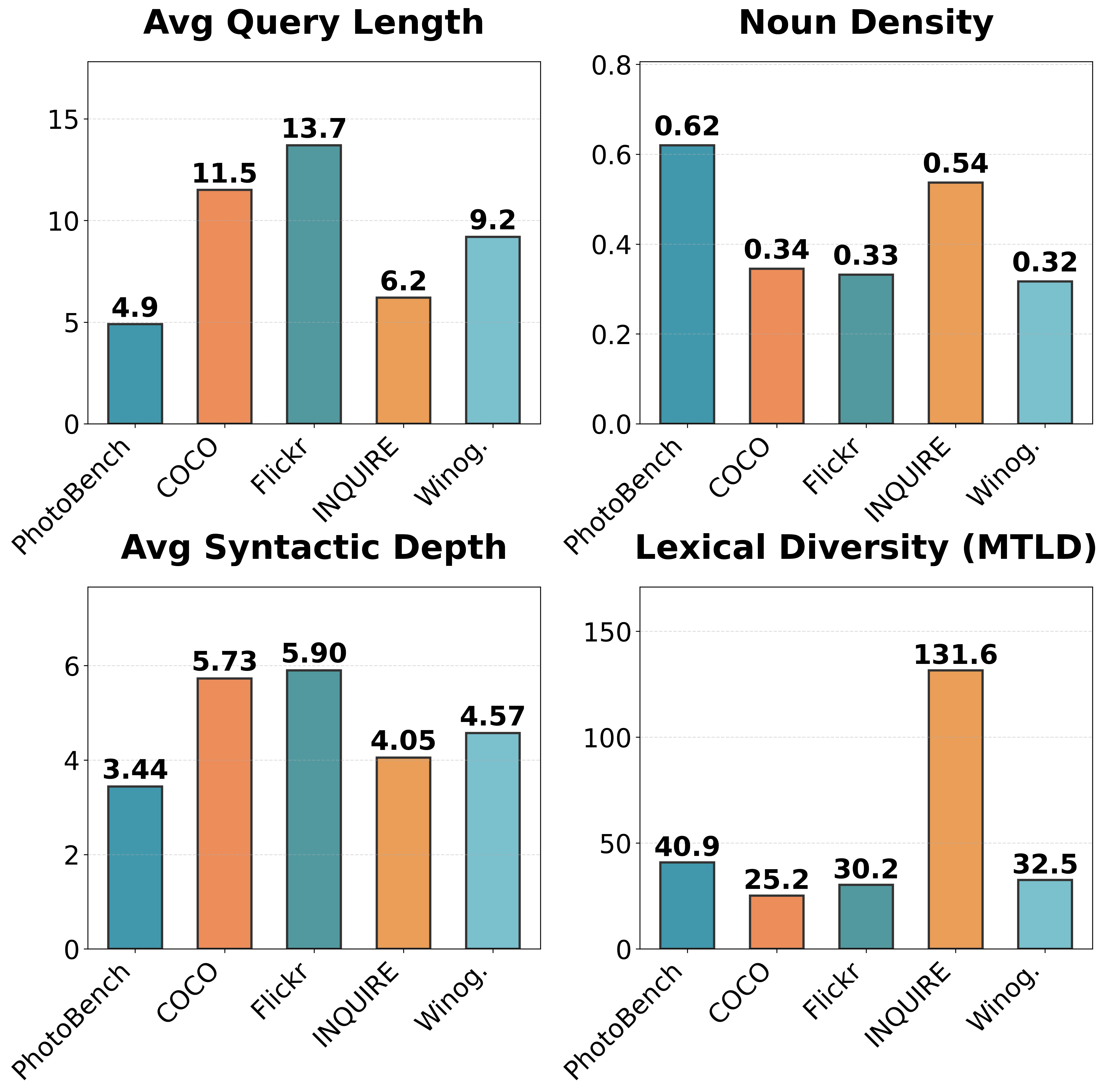}
    \caption{Linguistic comparison with traditional benchmarks. We evaluate four dimensions: Avg Query Length (total tokens), Noun Density (proportion of entities), Avg Syntactic Depth (grammatical complexity), and Lexical Diversity (MTLD, measuring vocabulary richness independent of text length). PhotoBench queries are structurally streamlined (lower length and depth) yet lexically diverse, reflecting search-style rather than narrative-style language.}
    \label{fig:dense_summary}
\end{figure}

As shown in Figure~\ref{fig:dense_summary}, \benchmark{} exhibits a linguistic profile distinct from traditional narrative benchmarks. To quantify the transition from "descriptive narration" to "search-style" retrieval, we evaluate queries across two dimensions: structural complexity (Length and Syntactic Depth) and semantic concentration (Noun Density and MTLD).

Unlike the verbose, clause-heavy captions in COCO2014 and Flickr30k, our queries are characterized by structural conciseness---favoring short, noun-phrase-centric inputs with significantly lower syntactic depth. This efficiency-first approach reflects how users prioritize intent over grammatical ornamentation. Crucially, this brevity does not equate to semantic simplicity; \benchmark{} consistently outperforms standard benchmarks in Lexical Diversity (MTLD), indicating a more specialized and diverse vocabulary tailored to personalized contexts. Technical formulations for these metrics are detailed in Appendix~\ref{appendix:lingui_metric}

\subsection{Per-Album Statistics}

Table~\ref{tab:album_stats_app} provides detailed statistics for each album in \benchmark{}.

\begin{table}[h]
\centering
\caption{Per-album statistics. Albums vary in social density (Core Faces) and visual patterns (Portrait ratio), ensuring benchmark diversity.}
\label{tab:album_stats_app}
\footnotesize
\renewcommand{\arraystretch}{1.2}
\begin{tabular}{lcccc}
\toprule
\textbf{Album} & \textbf{\# Images} & \textbf{Metadata (\%)} & \textbf{Portrait (\%)} & \textbf{Core Faces} \\
\midrule
Album 1 & 1,069 & 97.7 & 7.3 & 10 \\
Album 2 & 1,466 & 75.7 & 43.5 & 2 \\
Album 3 & 1,047 & 76.7 & 24.6 & 8 \\
\midrule
\textbf{Total} & \textbf{3582} & 83.4 & 25.1 & 20 \\
\bottomrule
\end{tabular}
\end{table}

\noindent\textbf{Album Diversity.} The three albums represent distinct user archetypes:
\begin{itemize}[leftmargin=*, itemsep=0pt]
    \item \textbf{Album 1} (event-centric): High metadata coverage (97.7\%), low portrait ratio (7.3\%), many social connections (10 core faces). Typical of users who document activities and locations.
    \item \textbf{Album 2} (person-centric): Highest portrait ratio (43.5\%), fewer core faces (2). Typical of users focused on family/couple documentation.
    \item \textbf{Album 3} (balanced): Moderate across all dimensions. Representative of general-purpose personal albums.
\end{itemize}

\subsection{Temporal Distribution}
\label{temperal_distribution}
\begin{figure}[h]
    \centering
    \includegraphics[width=0.95\columnwidth]{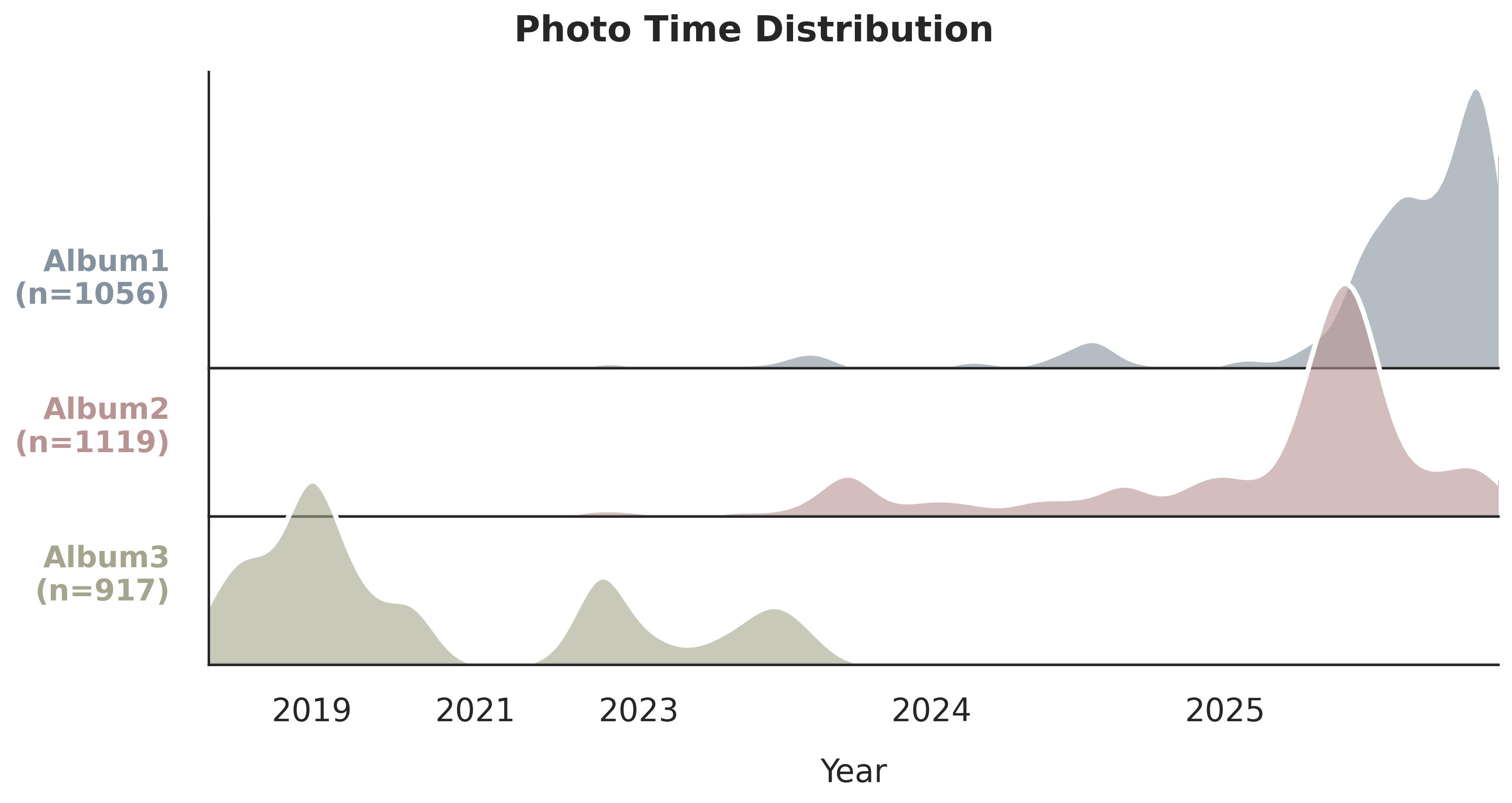}
    \caption{Temporal distribution of photos across albums.}
    \label{fig:time_zone_density_app}
\end{figure}

Figure~\ref{fig:time_zone_density_app} illustrates the temporal span of our benchmark, covering a diverse period from 2018 to 2025. Notably, the peak activities across the three albums are staggered and complementary. When one album exhibits a sparsity of data, another often compensates with a high density of photos. This interleaved distribution ensures a relatively uniform aggregate density over an extensive timeline, preventing temporal bias while providing a robust testbed for evaluating a model's long-range temporal reasoning and retrieval consistency.

\section{Supplementary Analysis: The Fact vs.\ Cognitive Dimension} \label{app:cognitive}
In this section, we introduce an exploratory taxonomy and its corresponding experiments. Despite the compelling theoretical motivation, current model bottlenecks prevent the results from fully manifesting the expected performance distinctions at this stage.
\subsection{Motivation and Taxonomy}
Initially, we hypothesized that the primary bottleneck in personal photo retrieval lies in the \textbf{reasoning depth} required by a query. To investigate this, we developed a dual-layer taxonomy focusing on five semantic dimensions: \textit{Location}, \textit{Time}, \textit{Person}, \textit{Object}, and \textit{Concept}. 

As defined in Table~\ref{tab:taxonomy_refined}, each dimension in a query is annotated as either \textbf{Fact-based} (requiring direct perceptual mapping) or \textbf{Cognitive} (requiring inferential reasoning, world knowledge, or personal context). Crucially, within a single dimension, these labels are mutually exclusive, though a single query often spans multiple dimensions (e.g., ``photos with my girlfriend at home'' involves Cognitive labels in both \textit{Person} and \textit{Location} dimensions).

\begin{table}[h]
\centering
\caption{The Dual-Layer Taxonomy across five dimensions. Fact-based queries involve explicit signal matching; Cognitive queries require reasoning synthesis.}
\label{tab:taxonomy_refined}
\small
\renewcommand{\arraystretch}{1.2}
\begin{tabularx}{\columnwidth}{lXX}
\toprule
\textbf{Dimension} & \textbf{Fact-based (Explicit)} & \textbf{Cognitive (Implicit)} \\ \hline
\textbf{Location} & ``in Shanghai'' & ``at home,'' ``at the gym'' \\
\textbf{Time} & ``in May 2024'' & ``during Spring Festival'' \\
\textbf{Person} & ``Zhang Wei'' & ``with my girlfriend'' \\
\textbf{Object} & ``a red car'' & ``the parking spot before the concert'' \\
\textbf{Concept} & ``sunset'' & ``cozy moments with family'' \\
\bottomrule
\end{tabularx}
\end{table}

\subsection{Statistical Distribution}
\begin{figure}[h]
    \centering
    \begin{subfigure}{0.48\linewidth}
        \centering
        \includegraphics[width=\linewidth]{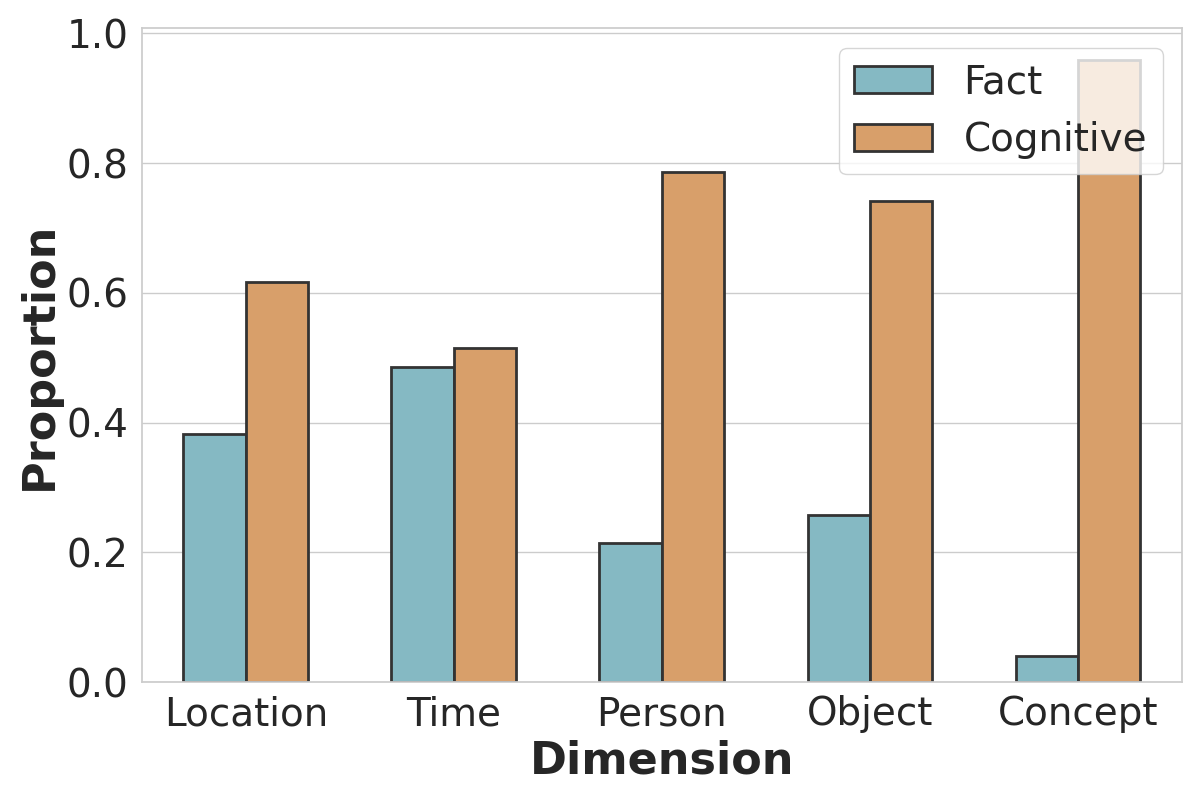}
        \caption{Cognitive vs.\ Fact Ratio}
        \label{fig:app_cog_fact}
    \end{subfigure}
    \hfill
    \begin{subfigure}{0.48\linewidth}
        \centering
        \includegraphics[width=\linewidth]{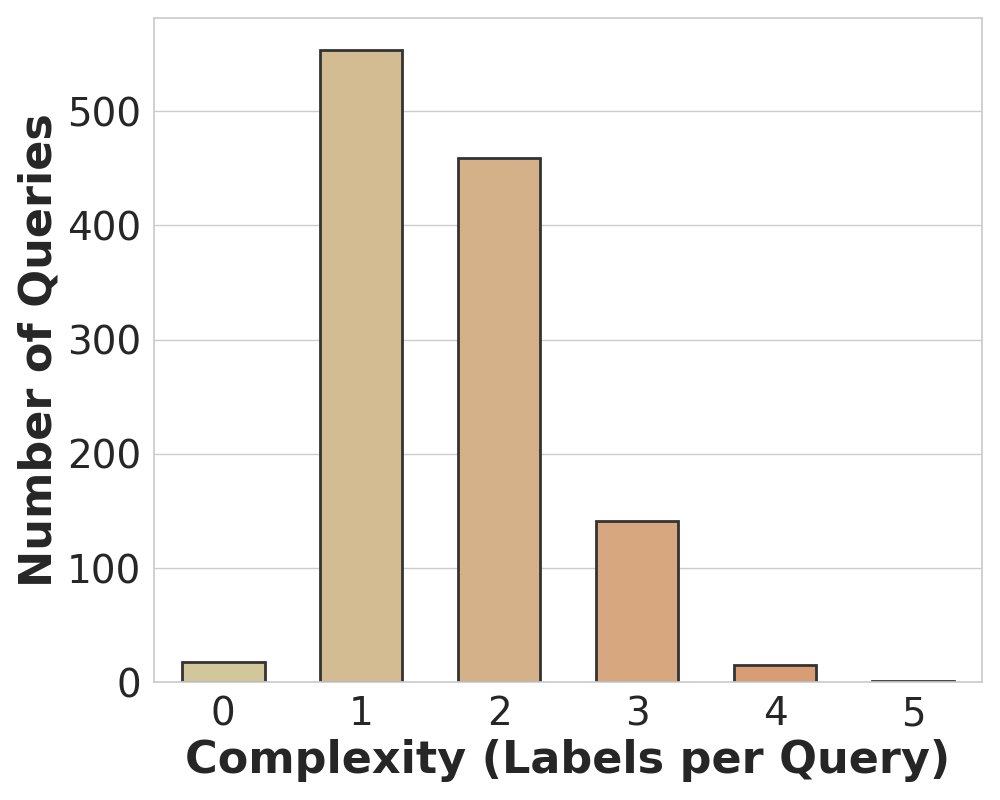}
        \caption{Cognitive Dimension Count}
        \label{fig:app_complexity}
    \end{subfigure}
    \caption{Supplementary query statistics. (a) Distribution of labels across semantic dimensions. (b) Distribution of Cognitive labels per query (0 indicates a pure Fact-based query).}
    \label{fig:app_stats}
\end{figure}

We annotated our entire dataset using this framework. Figure~\ref{fig:app_stats} illustrates the distribution:
\begin{itemize}[leftmargin=*, itemsep=0pt]
    \item \textbf{Dimension Ratio (a):} Shows the prevalence of Cognitive vs.\ Fact labels across dimensions. Cognitive requirements are most prominent in \textit{Time} and \textit{Concept} dimensions.
    \item \textbf{Cognitive Complexity (b):} Shows the number of Cognitive labels per query. A value of ``0'' denotes a purely Fact-based query or one with no specific dimension tags. Most queries involve 1--2 cognitive dimensions, confirming the inferential nature of real-world photo searches.
\end{itemize}

\subsection{The Counterintuitive Result}
Our analysis revealed a surprising trend: the ``Cognitive Gap'' is not the primary performance driver. As shown in Table~\ref{tab:fact_cognitive_app}, embedding models actually perform \textit{better} on Cognitive queries than on Fact queries ($-28.52$ gap).

\begin{table}[h]
\centering
\caption{Performance by reasoning depth (Recall@10). Embedding models show a negative gap (better on Cognitive), while agents show a slight positive gap.}
\label{tab:fact_cognitive_app}
\small
\setlength{\tabcolsep}{12pt}
\begin{tabular}{lccc}
\toprule
\textbf{Method Type} & \textbf{Fact} & \textbf{Cognitive} & \textbf{Gap} \\ \midrule
\textbf{Best Embedding} & 35.24 & 63.76 & -28.52 \\
\textbf{Best Agent}     & 78.60 & 70.98 & +7.62 \\
\bottomrule
\end{tabular}
\end{table}

\noindent\textbf{Interpretation: Modality Gap Masks Reasoning Gap.} 
For embedding models, the lower Fact performance is largely a byproduct of the \textbf{Modality Gap}: many Fact-based queries are Metadata-dependent (e.g., specific dates or locations), which are invisible to visual embeddings. Conversely, many Cognitive queries (e.g., ``cozy moments'') possess strong visual signatures that embeddings can capture through semantic similarity, even without explicit reasoning.

For agents, Cognitive queries are slightly more challenging due to the \textbf{Source Fusion Paradox}: resolving multiple implicit dimensions requires complex tool-calling and information pruning, which increases the likelihood of error propagation.

\subsection{Conclusion}
While the Fact/Cognitive distinction provides an intuitive lens, our empirical data suggests that \textbf{information-source accessibility (VMF)} is a more actionable diagnostic axis. We therefore prioritize the VMF taxonomy in our main analysis.


\section{Additional Experiments on English Version of \benchmark{}}
This part presents additional experiments on the English version of \benchmark{}. Overall, the experimental results are consistent with our primary findings. This demonstrates the generalizability of our benchmark and confirms that the observations are language-independent. Table \ref{tab:modality_gap_en} is the English version of Table ~\ref{tab:modality_gap}.

\begin{table}[h]
\centering
\caption{Recall@10 performance decomposition by source-aware query types (English Version).}
\label{tab:modality_gap_en}
\resizebox{0.47\textwidth}{!}{
\renewcommand\arraystretch{1.1}

\begin{tabular}{c|ccc|cc}
\toprule
\hline
Type & Agent(A) & Multimodal(M) & Text(T) & $\Delta$(A-M) & $\Delta$(A-T) \\
\hline
$S_V$    & 71.5 & 77.3 & 73.3 & -5.8  & -1.9  \\
$S_M$    & 53.4 & 6.5  & 7.8  & +46.9 & +45.6 \\
$S_F$    & 69.4 & 12.4 & 4.9  & +57.1 & +64.5 \\
$S_{VM}$ & 62.0 & 65.8 & 63.4 & -3.8  & -1.3  \\
$S_{VF}$ & 42.6 & 57.8 & 47.5 & -15.1 & -4.9  \\
$S_{MF}$ & 57.0 & 13.0 & 8.3  & +44.0 & +48.7 \\
$S_{VMF}$ & 39.4 & 54.8 & 43.9 & -15.4 & -4.5  \\
\hline
\bottomrule

\end{tabular}}
\end{table}


\section{Experimental Protocol of Commercial System}
\label{app:phone_detail}

For commercial system evaluation, we followed this standardized procedure:
\begin{enumerate}[leftmargin=*, itemsep=0pt]
    \item Factory reset each device to ensure clean state.
    \item Transfer all album images via USB.
    \item Allow 24 hours for on-device indexing and automatic organization to complete.
    \item Issue queries via native gallery search interface.
    \item Record all returned results (up to 100 per query).
    \item Two annotators independently verified result relevance.
\end{enumerate}

\section{Case Study Gallery}
\label{app:cases}

This section presents additional case studies illustrating the contrast between embedding-based and agentic retrieval.

\subsection{Case: Functional Intent}

\textbf{Query:} ``The receipt I used for reimbursement on my last business trip''

\textbf{Analysis:} This query requires: (1) identifying ``business trip'' events from trajectory logs, (2) locating the most recent occurrence, and (3) finding receipt-like images within that specific temporal window.

\textbf{Embedding Result:} Returns visually similar receipt images from various time periods, including unrelated personal purchases, due to a lack of temporal and logical constraints.

\textbf{Agent Result:} Correctly identifies the specific business trip event (e.g., 3 weeks ago), filters the search to that time range, and then isolates receipt images, successfully returning the correct documentation.

\subsection{Case: Social Context}

\textbf{Query:} ``Photos of New Year's Eve dinner with my parents''

\textbf{Analysis:} This requires: (1) resolving ``parents'' to specific face IDs via identity metadata, (2) identifying the specific Spring Festival timeframe, and (3) locating dining-related images containing those specific individuals.

\textbf{Embedding Result:} Returns generic dinner scenes and family photos without temporal specificity, failing to distinguish the holiday context.

\textbf{Agent Result:} Utilizes \texttt{identity\_lookup} for parents, filters for the Lunar New Year period, and identifies dining scenes, correctly retrieving the specific celebration photos.

\subsection{Case: False Memory (Zero-GT)}
\label{zeroGT}

\textbf{Query:} ``Sunset photos at the beach last summer''

\textbf{Ground Truth:} No beach photos exist in the user's album for the specified period.

\textbf{Embedding Result:} Returns sunset photos from other locations and beach photos from different years---yielding false positives due to partial visual matching.

\textbf{Agent Result:} Executes a filtered search for beach locations within the summer timeframe, finds no matches, and correctly returns an empty result set with a logical explanation.

\subsection{Case: Query types}
\label{app:query_case_study}

\begin{figure}[t]
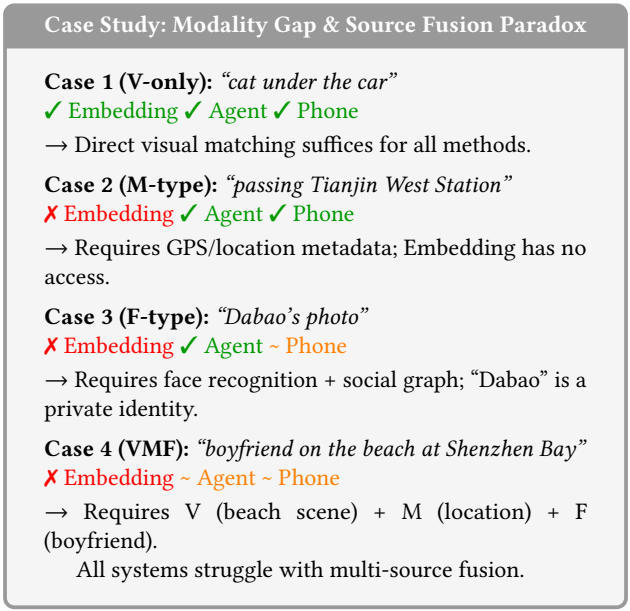

\centering
\begin{tcolorbox}[colback=gray!8, colframe=gray!80, title=\textbf{Case Study: Modality Gap \& Source Fusion Paradox}, width=0.98\columnwidth, coltitle=white, fonttitle=\bfseries]
\textbf{Case 1 (V-only):} \textit{``cat under the car''}\\
\textcolor{green!60!black}{\ding{51} Embedding} \textcolor{green!60!black}{\ding{51} Agent} \textcolor{green!60!black}{\ding{51} Phone}\\[2pt]
$\rightarrow$ Direct visual matching suffices for all methods.

\vspace{4pt}
\textbf{Case 2 (M-type):} \textit{``passing Tianjin West Station''}\\
\textcolor{red}{\ding{55} Embedding} \textcolor{green!60!black}{\ding{51} Agent} \textcolor{green!60!black}{\ding{51} Phone}\\[2pt]
$\rightarrow$ Requires GPS/location metadata; Embedding has no access.

\vspace{4pt}
\textbf{Case 3 (F-type):} \textit{``Dabao's photo''}\\
\textcolor{red}{\ding{55} Embedding} \textcolor{green!60!black}{\ding{51} Agent} \textcolor{orange}{\textasciitilde~Phone}\\[2pt]
$\rightarrow$ Requires face recognition + social graph; ``Dabao'' is a private identity.

\vspace{4pt}
\textbf{Case 4 (VMF):} \textit{``boyfriend on the beach at Shenzhen Bay''}\\
\textcolor{red}{\ding{55} Embedding} \textcolor{orange}{\textasciitilde~Agent} \textcolor{orange}{\textasciitilde~Phone}\\[2pt]
$\rightarrow$ Requires V (beach scene) + M (location) + F (boyfriend).\\
\hspace*{12pt}All systems struggle with multi-source fusion.
\end{tcolorbox}
\caption{Case study illustrating the Modality Gap and Source Fusion Paradox. \textcolor{green!60!black}{\ding{51}}=success, \textcolor{orange}{\textasciitilde}=partial, \textcolor{red}{\ding{55}}=failure.}
\label{fig:case_study}
\end{figure}

Figure~\ref{fig:case_study} illustrates representative query cases across different source types, as well as the performance of different methods:

\noindent\textbf{Case 1 (V-only):} Query: ``cat under the car''. All approaches succeed---direct visual matching suffices.

\noindent\textbf{Case 2 (M-type):} Query: ``passing Tianjin West Station''. Embedding fails completely (no location reasoning); Agent and Phone succeed via metadata filtering.

\noindent\textbf{Case 3 (F-type):} Query: ``Dabao's photo''. Embedding fails as it cannot resolve private identities. Agent succeeds via Face Engine; Phone performance varies by vendor.

\noindent\textbf{Case 4 (VMF):} Query: ``boyfriend on the beach at Shenzhen Bay''. This requires visual scene understanding, location/time metadata, and face recognition simultaneously. All systems show degraded performance, demonstrating the Source Fusion Paradox.

\section{Agentic Retrieval Framework}
\label{app:framework_results}

While the main text evaluates a general-purpose agentic baseline for benchmarking, this section details a specialized, multi-stage Agentic Retrieval Framework specifically architected for the nuances of personal photo management, which is illustrated in Fig.~\ref{fig:ap_agentic_framework}. Compared to the standard agent, this framework introduces a more sophisticated orchestration layer to handle the semantic complexity of gallery-based queries.

As illustrated in Fig.~\ref{fig:ap_agentic_framework}, the architecture leverages a unified 4B Vision-Language Model to function as the internal Planner, Evaluator, and captioning engine, ensuring cross-task consistency. The underlying retrieval stage is powered by a high-efficiency 2B VLM-based embedding model.

The framework adopts a three-phase routing mechanism: it first executes rule-based matching, followed by intent-driven distribution to either metadata lookups or hybrid retrieval. For complex queries requiring deeper reasoning, the system escalates to the agentic planner to synthesize results. Preliminary results in Table~\ref{tab:agentic_metrics} demonstrate that our framework achieves a superior F1 score of 62.07\%, outperforming existing baselines. These findings suggest that optimizing agentic coordination is a promising direction for enhancing retrieval performance in on-device mobile environments.

\section{Agentic Retrieval Framework}
\label{app:framework_results}

\begin{figure}[h]
    \centering
    \includegraphics[width=1.0\columnwidth]{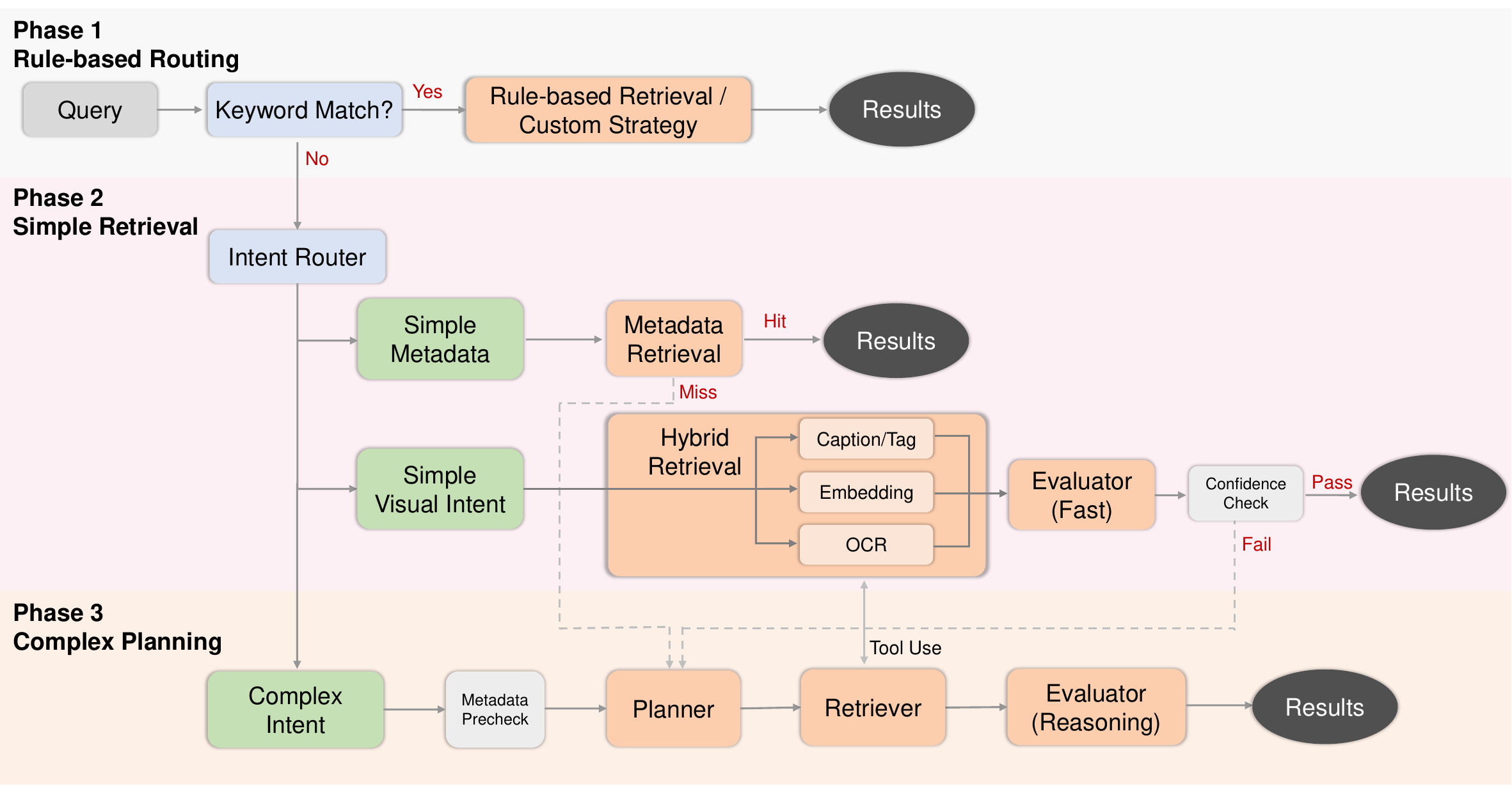}
    \caption{The proposed hierarchical agentic framework architecture for photo retrieval.}
    \label{fig:ap_agentic_framework}
\end{figure}

While the main text evaluates a general-purpose agentic baseline for benchmarking, this section details a specialized, multi-stage Agentic Retrieval Framework (illustrated in Fig.~\ref{fig:ap_agentic_framework}) specifically architected for the nuances of personal photo management. Compared to the standard agent, this framework introduces a more sophisticated orchestration layer to handle the semantic complexity of gallery-based queries.

The architecture leverages a unified 4B Vision-Language Model (VLM) to function as the Planner, Evaluator, and captioning engine, while the underlying retrieval stage is powered by a 2B VLM-based embedding model. Orchestrating these components is a three-phase routing mechanism that initiates with rule-based matching, progresses to intent-driven distribution—directing queries to either metadata lookups or hybrid retrieval—and ultimately escalates complex requests requiring deeper reasoning to the agentic planner for comprehensive result synthesis.

As shown in Table~\ref{tab:ours_metrics}, our framework achieves the best overall performance, with the highest F1 score of 63.3\% on normal queries and the strongest rejection capability (Rej-F1 of 52.0\%). While our recall (70.2\%) is slightly lower than the best agentic baseline, this is primarily due to our significantly smaller retrieval model (2B parameters) compared to large-scale models. These findings suggest that optimizing agentic coordination with lightweight on-device models is a promising direction for enhancing retrieval performance in mobile environments.

\begin{table}[h]
\centering
\caption{Performance of our proposed agentic framework on \benchmark{}. All values in \%. ``\textit{P}", ``\textit{R}", ``\textit{Rej-P}", ``\textit{Rej-R}" and ``\textit{Rej-F1}" denote Precision, Recall, F1, Reject-Precision, Reject-Recall and Reject-F1, respectively.}
\label{tab:ours_metrics}
\vspace{-8pt}
\footnotesize
\resizebox{0.47\textwidth}{!}{
\renewcommand\arraystretch{1.1}
\begin{tabular}{c|ccc|ccc}
\toprule
\hline
\multirow{2}{*}{System} & \multicolumn{3}{c|}{Normal Query} & \multicolumn{3}{c}{Zero-GT Query} \\
  & P   & R  & F1  & Rej-P   & Rej-R  & Rej-F1   \\ 
  \hline
Ours & \textbf{67.3} & 70.2 & \textbf{63.3} & \textbf{50.8} & \textbf{53.3} & \textbf{52.0} \\
ToolACE-2               & 38.3         & 54.7     & 39.0   & 15.1     & \underline{35.0}     & 21.1     \\
Qwen3-8B                & 37.6         & 68.1     & 41.0   & 16.8     & 16.7     & 16.7     \\
Qwen3-32B               & 39.6         & 66.3     & 41.4   & 16.7     & 25.8     & 20.3     \\
DeepSeek-v3             & 38.5         & 66.2     & 40.7   & 11.8     & 10.0     & 10.8     \\
Qwen3-235B-A22B         & \underline{50.8} & 71.6 & \underline{51.9} & 14.3 & 6.7 & 9.1 \\
GPT-4o                  & 45.4         & 65.3     & 46.4   & 13.8     & 17.5     & 15.4     \\
OpenAI-o3               & 44.6         & \underline{75.2}     & 48.0   & \underline{34.5}     & 33.3     & \underline{33.9}     \\
Claude-Sonnet-4.5       & 43.7         & \textbf{78.3}     & 47.9   & 14.3     & 8.3      & 10.5     \\
Claude-Opus-4.5         & 46.9         & 74.7     & 50.5   & 18.2     & 13.3     & 15.4     \\
\hline
\bottomrule
\end{tabular}}
\end{table}





\section{Statement on Evaluation Discrepancies}
Current commercial mobile retrieval systems are often highly optimized for specific visual perception scenarios (e.g., identity documents, visual objects, or pets) using specialized models and hard-coded metadata rules. We clarify that our benchmark is not intended to provide an exhaustive evaluation of every fine-grained visual category. Furthermore, due to the difference in evaluation dimensions and data distributions, it is both expected and reasonable that our results may diverge from phone manufacturers' internal benchmarks. Our primary objective is to evaluate the system’s capability to holistically satisfy complex and deep-seated retrieval intents, moving beyond simple recognition to deep semantic understanding.

\end{document}